\documentclass[lettersize, journal,colorlinks=true, linkcolor=blue, citecolor=blue, urlcolor=blue]{IEEEtran}
\usepackage{amsmath,amsfonts}
\usepackage{algorithmic}
\usepackage{algorithm}
\usepackage{array}
\usepackage[font=footnotesize]{caption}
\usepackage{subcaption}
\usepackage{textcomp}
\usepackage{stfloats}
\usepackage{url}
\usepackage{verbatim}
\usepackage{graphicx}
\usepackage{booktabs}
\usepackage{orcidlink}
\usepackage{cite}
\usepackage{setspace}
\usepackage{xcolor}
\usepackage{verbatim}
\usepackage{enumitem}

\begin{document}

\bstctlcite{IEEEexample:BSTcontrol}

\title{Physics-Informed PointNets for Modeling Electromagnetic Scattering from All-Dielectric Metasurfaces with Inclined Nanopillars}

\author{Leon Armbruster \orcidlink{0009-0007-7602-362X}, 
        Vlad Medvedev \orcidlink{0009-0007-2445-2617} and
        Andreas Rosskopf \orcidlink{0000-0003-1854-1297}%
\thanks{(Corresponding author: Leon Armbruster)
The authors are with the Fraunhofer Institute for Integrated Systems and Device Technology, 91058 Erlangen, Germany (e-mail: leon.armbruster@iisb.fraunhofer.de; vlad.medvedev@iisb.fraunhofer.de; andreas.rosskopf@iisb.
fraunhofer.de).}%
\thanks{Manuscript received Month Day, Year; revised Month Day, Year.}}



\maketitle
\begin{abstract}
    \small
    Metasurfaces are innovative planar optical structures capable of manipulating incident light properties. Accurate and computationally efficient modeling of such metasurfaces, particularly those with irregular geometries, remains a challenge for conventional solvers. In this work, we present a mesh-free Physics-Informed PointNet (PIPN) to model electromagnetic scattering from all-dielectric metasurfaces that feature spatially varying nanopillars. Our approach uses the PointNet architecture to directly encode spatially varying material properties into the Physics-Informed Machine Learning (PIML) framework. We demonstrate the generalization capability of our PIPN through evaluations on datasets; these datasets are generated with varying refractive indices representing common dielectric materials. Furthermore, the inclination angles are varied within each dataset, which represent expected manufacturing defects.
\end{abstract}



\section{Introduction}
\IEEEPARstart{R}{apid} advancements in nanofabrication techniques have led to the development of sophisticated nano-optic devices, with features on the scale of (or even smaller than) the dimensions of the operating wavelength. In particular, metasurfaces have emerged as planar optical structures capable of manipulating incident light in diverse ways.\\

Metasurfaces are artificially engineered planar materials composed of meta-atoms with subwavelength dimensions. Because they allow precise control over the properties of the electromagnetic waves they interact with, metasurfaces are an important advancement in the field of nano-optics~\cite{neshev_optical_2018, kuznetsov2024roadmap, yu2013flat, qin2022metasurface}. By scattering light from resonant nanostructures, properties-such as phase, amplitude, and polarization-can be finely tuned, as opposed to conventional optical elements that reflect, scatter, or refract light in a less controlled manner. This unlocks novel applications in fields such as: biomedical imaging and sensing~\cite{kim_optical_2025}, full color holography~\cite{hu_3d-integrated_2019}, multiplexed communications~\cite{nie_metasurfaces_2021}, and all-optical diffractive deep neural networks~\cite{liu_programmable_2022}.

Due to these new applications, demand has increased for accurate and efficient modeling techniques to predict the electromagnetic behavior of metasurfaces and optimize their performance. To model light diffraction from these sub-wavelength nanostructures and capture complex optical effects, researchers have traditionally relied on electromagnetic field solvers. These traditional solvers numerically approximate Maxwell's equations by utilizing algorithms such as the finite-difference/element methods~\cite{teixeira_time-domain_2008} and rigorous coupled wave analysis~\cite{evanschitzky_fast_2007}. However, difficulties can arise when the dimensionality of the problem increases or the governing equations involve highly nonlinear or multi-scale behavior. These limitations are exacerbated when simulating curvilinear or irregular nano-optic geometries, where a high degree of spatial discretization is required to resolve the interactions between light and the nanoscale features~\cite{kang_large-scale_2024}.\\
\IEEEpubidadjcol 
These flaws have led to the development of alternative approaches capable of maintaining the high accuracy numerical solvers provide while reducing computational time and complexity. Recent advancements in machine learning---particularly deep learning---have introduced a new paradigm to address challenges in modeling of nano-optical devices. Early deep-learning applications in computational optics have been shown to deliver faster inference once trained and strong generalization to unseen data. These studies demonstrate Neural Networks' (NNs) ability to predict electromagnetic responses~\cite{an_deep_2022, li_pisc-net_2024} (such as far-field spectra) with high accuracy. However, these data-driven approaches often require extensive training datasets generated through costly simulations or obtained by experimental measurements and long training times. Moreover, they operate as 'black-box' models, offering limited insight into the underlying physics and often struggling with interpolation or extrapolation tasks. To overcome these limitations, hybrid approaches~\cite{cuomo_scientific_2022} have gained traction, combining data-driven methods with physics-based constraints. Physics-Informed Machine learning (PIML) has emerged as a prominent extension of traditional data-driven machine learning approaches~\cite{raissi_physics_2017, toscano_pinns_2025, tu_physics-informed_2023, nguyen_physics-informed_2024, cuomo_scientific_2022}. So-called Physics-Informed Neural Networks (PINNs) incorporate the governing equations of physical systems directly into the NN training process, constraining the network to produce physically plausible solutions, thereby reducing overfitting to the provided data and improving interpretability.

\begin{figure}[!ht]
    \centering
    \includegraphics[width=3.5in]{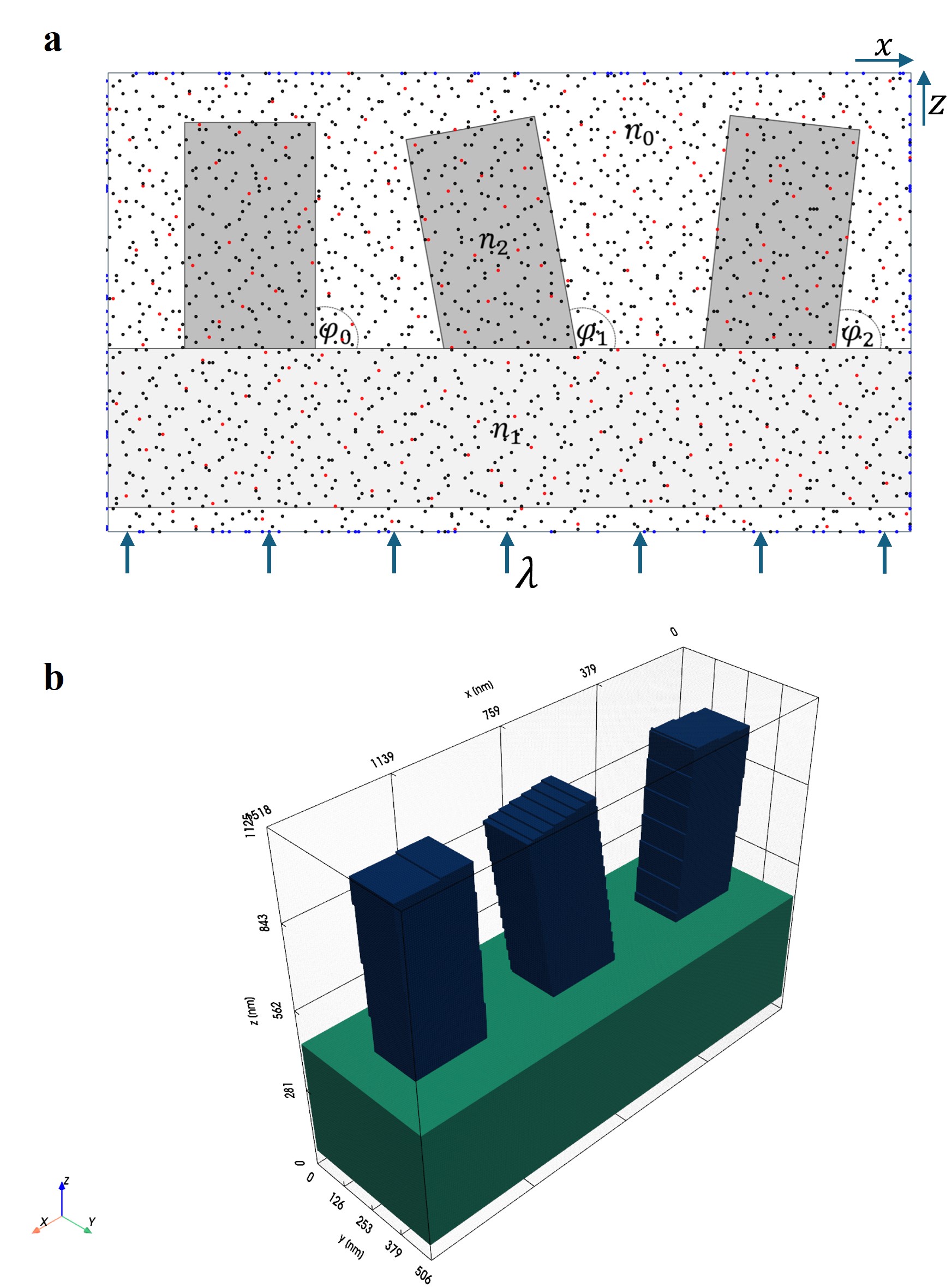} 
    \caption{A representative schematic of a (a) 2D and (b) 3D metasurface domain where $n_0, n_1 \,\text{and} \,n_2$ are the refractive indices for the different materials, $(\varphi_0, \varphi_1, \varphi_2)$ are the inclination angles for the pillars, and $\mathbf{\lambda}$ is wavelength of the source plane wave. In (a) a representative training point cloud ($N_t$=2048) is drawn with boundary (blue), interior (black), and reference (red) points.}
    \label{fig:example_domain}

\end{figure}
In addition, PINNs can incorporate observational data directly into the loss function, allowing for additional constraints in settings where data-driven regularization is necessary~\cite{gopakumar_loss_2023}. This can be especially useful when sparse observational or experimental data is available, or when the underlying physics constraints are not exactly known. Despite these advantages, PINNs introduce their own set of challenges, such as slow convergence in training, difficulty in balancing different loss terms, and sensitivity to hyperparameter tuning~\cite{wang_experts_2023, wang_understanding_2021}. 

Early applications of PINNs in nano-optics have shown promise in solving inverse scattering problems involving interacting nanostructures~\cite{10.1063/5.0071616, 10618982, Wang:24}, designing photonic devices~\cite{kojima_inverse_2023}, and lithography mask simulations~\cite{medvedev_3d_2024_1, medvedev_3d_2024,Medvedev:25}. However, many existing studies have relied on scalar approximations and mesh-based methods that limit their ability to efficiently capture electromagnetic interactions on complex domains. 

To overcome these limitations, we employed a Physics-Informed PointNet (PIPN) framework. Unlike mesh-based methods, PIPN frameworks use a point cloud representation of the domain, which enables efficient encoding of varying domain geometries. To the best of our knowledge, this is the first work that encodes spatially varying Partial Differential Equation (PDE) parameters into the PIPN training framework. Based on that, we can train NNs that learn the connection between the material properties of the metasurface and the corresponding electromagnetic response. This allows for subsequent fast inference on previously unseen domains. By extending the PIPN framework, we aim to develop a robust and scalable mesh-free method for predicting both near-field and far-field electromagnetic responses of nano-optical devices.

In \hyperref[sec:GoverningEquations]{Sec.~\ref*{sec:GoverningEquations}}, we describe the physical model we used as well as our reason for including reference solutions in our training paradigm. Then, in \hyperref[sec:PointNet]{Sec.~\ref*{sec:PointNet}} we introduce the PointNet architecture and highlight its advantage over mesh-based approaches. Next, we discuss in greater detail how the training and testing domains are constructed, and how the reference solutions are calculated. In \hyperref[sec:lossFunction]{Sec.~\ref*{sec:lossFunction}}, we construct the loss functions used during training to highlight the addition of a reference loss term.~\hyperref[sec:Implementation]{Sec.~\ref*{sec:Implementation}} specifies the training hyperparameters and discuss the metrics used for testing. The main results are presented in \hyperref[sec:Results]{Sec.~\ref*{sec:Results}} and discussed in \hyperref[sec:discussion]{Sec.~\ref*{sec:discussion}}; the paper concludes in \hyperref[sec:Conclusion]{Sec.~\ref*{sec:Conclusion}}.

\section{Problem Formulation}
\label{sec:ProblemFormulation}

\subsection{Governing Equation}
\label{sec:GoverningEquations}

The Helmholtz equation is widely used in scattering problems because it reduces steady-state wave propagation to a spatial boundary-value problem. Whereas Maxwell’s equations describe the full time-dependent behavior of electromagnetic fields, the Helmholtz equation follows from the time-harmonic assumption. Assuming a monochromatic light, we write \(\mathbf{E}(\mathbf{x}, t) = \mathbf{E}(\mathbf{x}) e^{-i\omega t}\), where \(\omega\) is the angular frequency and \(\lambda\) is the wavelength of the incident light. For linear, isotropic, non-magnetic, and non-dispersive media with no free current density, substituting this representation into Maxwell’s curl equations yields the frequency-domain vector Helmholtz equation:
\begin{equation}
    \label{eq:helmholtz}
    \nabla \times (\nabla \times  \mathbf{E}(\mathbf{x})) +  \omega^2 \varepsilon(\mathbf{x}) \mathbf{E}(\mathbf{x}) =  0,
\end{equation}
where \(\mathbf{x} = (x, z)\) for the 2D case and \(\mathbf{x} = (x, y, z)\) for the 3D case. We furthermore assume lossless media with no absorption or amplification effects, ensuring energy conservation by setting $\text{Im}(\varepsilon(\mathbf{x})) = 0$, where $\varepsilon(\mathbf{x})$ is the spatially dependent refractive index. In addition to~\eqref{eq:helmholtz}, we impose periodic boundary conditions in the $x$- and $y$-direction. In the $z$-direction, non-reflective boundary conditions are assumed. Adding the boundary conditions to~\eqref{eq:helmholtz} gives us the problem formulation of the Helmholtz equation for the 3D case.

For the 2D case, we can effectively decompose the 2D incident field into two primary polarization states: Transverse Electric (TE) and Transverse Magnetic modes. In our model, we have a TE-polarized monochromatic plane wave traveling in the $z$-direction. In the TE polarization, the electric field is entirely perpendicular to the plane of incidence ($xz$-plane). This implies that the electric field \(\mathbf{E}\) has only a single non-zero component, \(\mathbf{E}_y\), which oscillates along the \(y\)-axis. As a result, we obtain the following scalar formulation of the Helmholtz equation and boundary conditions:

\begin{equation}
    \label{eq:helmholtz_te_pol}
    \begin{aligned}
        \nabla^2 \mathbf{E}_y(\mathbf{x})  +  \omega^2 \varepsilon(\textbf{x})\mathbf{E}_y(\mathbf{x}) &= 0 \quad &\mathbf{x} \in \Omega \subset \mathbb{R}^2, \\
         \mathbf{E}_y(\mathbf{x}_{\Gamma_a}) - \mathbf{E}_y(\mathbf{x}_{\Gamma_b}) & = 0 \quad &\mathbf{x}_{\Gamma_{a/b}}  \in \partial\Omega \subset \mathbb{R}^{2},\\
         \nabla \mathbf{E}_y(\mathbf{x}_{\Gamma_a}) - \nabla\mathbf{E}_y(\mathbf{x}_{\Gamma_b}) & = 0,
    \end{aligned}
\end{equation}

where $\mathbf{x}_{\Gamma_{a/b}}$ denote corresponding mirrored points on the boundaries of the domain $\Omega$. This is the scalar Helmholtz equation for the TE polarization case. Well-posedness for problem~\eqref{eq:helmholtz_te_pol} in the sense of Hadamard is shown in~\cite{graham_helmholtz_2019}. We henceforth set \(\mathbf{E}(\mathbf{x}) := \mathbf{E}_y(\mathbf{x})\). 

Furthermore, since no source is defined, problem~\eqref{eq:helmholtz_te_pol} admits a trivial solution where $ \mathbf{E}(\mathbf{x}) = 0 $ everywhere in $ \Omega $. While mathematically valid, this solution is physically meaningless unless justified by specific boundary conditions or constraints. One way to circumvent this limitation is to define the incident field either by adding source term to~\eqref{eq:helmholtz_te_pol} or by decomposing $\mathbf{E}_y(\mathbf{x})$ in~\eqref{eq:helmholtz_te_pol} onto scattered and incident fields~\cite{lim_maxwellnet_2022}. Another is to augment the training point clouds with points where the reference solutions is known, explicitly incorporating the excitation into the formulation; this is the approach adopted in this work.

\subsection{Domain Geometry}
\label{sec:DomainGeometry}

Given \eqref{eq:helmholtz}, we employed a rigorous-coupled wave analysis solver based on the waveguide method~\cite{evanschitzky_fast_2007} to compute reference solutions. In this work, we simulate transmission-type metasurfaces that are illuminated by a monochromatic light with $\mathbf{\lambda} = 750 \, \text{nm}$. The plane wave is incident on the substrate, and subsequently propagates through the meta-atom layer. The domain size remains constant, with the $x$-dimension set to $1500 \, \text{nm}$ (simulated by three units cells each with a period of $500 \, \text{nm}$) and the $z$-dimension set to $1125 \,\text{nm}$. In the 3D case, the $y$-dimension is set to $500 \, \text{nm}$. The typical height and width for each of the pillars are $500 \, \text{nm}$ and $250 \, \text{nm}$, respectively. The basis configuration is then three pillars equally spaced between one another on top of a substrate. The spacing between the bottom of the domain and the substrate, as well as the substrate thickness, remain fixed across all simulations. Due to numerical inaccuracies, minor fluctuations in the pillar width and height may occur on a scale of tens of nanometers. The spatial resolution of our dataset corresponds to $ \frac{750}{120}=6.25$ nanometers per `pixel`. 2D and 3D representative domains can be seen in \hyperref[fig:example_domain]{Fig.~\ref{fig:example_domain}a} and \hyperref[fig:example_domain]{Fig.~\ref{fig:example_domain}b}, respectively. 
The metasurfaces are surrounded by air, which has a refractive index of $ n_0 \approx 1.0 $. For visible and near-infrared metasurface applications, dielectric materials such as silicon dioxide ($\mathrm{SiO_2}$) and titanium dioxide ($\mathrm{TiO_2}$) are often used because of their low loss at those wavelengths compared to metals. Thus, three different cases are considered: 

\begin{enumerate}[label=({\Roman*})]
    \setlength\itemsep{1em}
    \item For the 2D SiO\textsubscript{2} case both the pillars and substrate are made of SiO\textsubscript{2} with refractive index \(n_1 = n_2 = 1.45\) (see \hyperref[sec:low_refractive_index]{Sec. \ref{sec:low_refractive_index}-Case I}).
    \item The same as (Case I), but in three-dimensions (see \hyperref[sec:3d_homogeneous]{Sec. \ref{sec:3d_homogeneous}-Case II}).
    \item The 2D TiO\textsubscript{2}-on-SiO\textsubscript{2} metasurface case features TiO\textsubscript{2} pillars (\(n_2 = 2.53\)) on a SiO\textsubscript{2} substrate (\(n_1 = 1.45\)), corresponding to a practical high-index-contrast implementation (see \hyperref[sec:high_refractive_index]{Sec. \ref{sec:high_refractive_index}-Case III}).
\end{enumerate}

\vspace{3mm}

To generate the datasets for the different cases, all geometric parameters of the domains are held constant except for the inclination angles of the three pillars, represented by the angles $(\varphi_0, \varphi_1, \varphi_2)$. These inclination angles are varied randomly, allowing the generation of distinct dataset configurations. Each inclination angle ranges from $-10^\circ$ to $10^\circ$ away from the substrate normal (aligned with the $z$-axis); this simulates realistic fabrication imperfections.

Each dataset consists of a total of $ |\mathcal{D}| = 500 $ samples split apart into \(\mathcal{D}_t \cup \mathcal{D}_v\) with training set $ \mathcal{D}_t $ of size $ 450 $ and a testing set $ \mathcal{D}_v $ of size $ 50 $.

\section{Methodology}
\label{sec:Methodology}

\subsection{PointNets}
\label{sec:PointNet}

\begin{figure*}[!ht]
    \centering
    \includegraphics[width=7in]{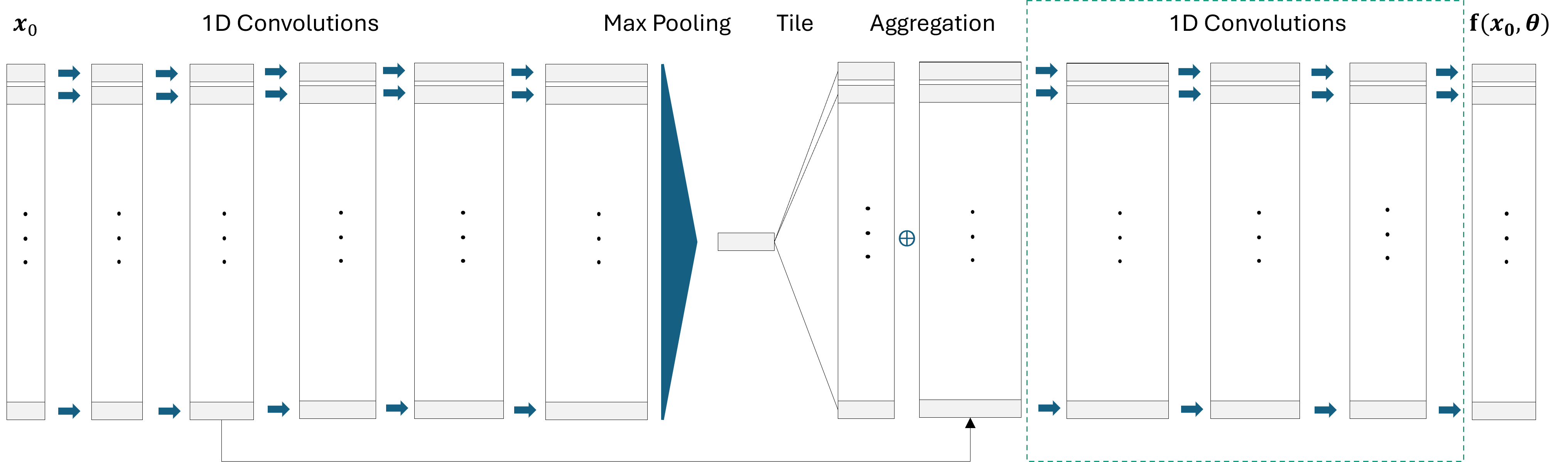}
    \caption{A schematic of the PointNet architecture used in this work. The PointNet's design leverages 1D convolutional kernels applied to each point individually to process local features, as well as a symmetric aggregation function (such as max pooling or average pooling) to generate a global feature vector representation that is invariant to the input order. These global and local feature vectors are combined to produce the output point cloud. The dotted square represents the second half of convolutional layer which can be replaced by other layer types (\hyperref[app:varying_pointnet_architectures]{App.~\ref{app:varying_pointnet_architectures}}).}
    \label{fig:pointnetc_schematic}
\end{figure*}

The PointNet (\hyperref[fig:pointnetc_schematic]{Fig.~\ref{fig:pointnetc_schematic}}) is a NN architecture originally developed for processing unordered point clouds for tasks in computer vision, including object classification \cite{garcia-garcia_pointnet_2016}, and segmentation \cite{qi_pointnet_2017}. This architecture was extended for use in PIML by \cite{kashefi_physics-informed_2022, kashefi_physics-informed_2023, kashefi_kolmogorov-arnold_2024}, which introduced and popularized PIPNs. The schematic in \hyperref[fig:pointnetc_schematic]{Fig.~\ref{fig:pointnetc_schematic}} illustrates how the PointNet processes its input.

Our input point cloud is structured as tensor $\mathbf{x}_0$ of shape $(N_t\times(d+1))$, where $N_t$ denotes the total number of points in the point cloud. The remaining dimension combines the spatial coordinates (determined by dimension $d$) with the spatially dependent permittivity. For the 2D problem, the network produces an output point cloud $\mathbf{f}(\mathbf{x}_0;\theta)$ of shape $(N_t\times2)$, where the final channel contains the real and imaginary parts of the electric field at each point of the input point cloud. In the 3D case, the output expands to $(N_t\times6)$, capturing the real and imaginary components of all three electric field vectors ($E_x$, $E_y$, and $E_z$). 

A modification to this network structure investigated in \hyperref[app:varying_pointnet_architectures]{App.~\ref{app:varying_pointnet_architectures}} is to replace the 1D convolutional layers after the aggregation (as indicated by the dashed square in \hyperref[fig:pointnetc_schematic]{Fig.~\ref{fig:pointnetc_schematic}}) either with dense multilayer perceptron layers or a Kolmogorov-Arnold Network (KAN)~\cite{liu_kan_2024}. This is viable because the material distribution and geometry have already been encoded after the feature aggregation. PIPNs offer several distinct advantages compared to mesh-based physics-informed NNs, which typically process binary images where pixels encode different regions (e.g., solid objects and fluid flow~\cite{ZHANG2022110179}).

Mesh-free approaches have the following advantages: First, for mesh-based NNs to capture the high frequency aspects of the input and output, the overall node density of the grid must be increased throughout the entire computational domain. This leads to an increase in computational time. A point cloud based representation offers a straight-forward way to overcome this problem, as sampling more points in regions where finer detail is needed allows for adaptive resolution since one can simply sample more points in regions where finer detail is needed to allow for adaptive resolution.

Second, defining boundary conditions can be challenging when working with irregular geometries in mesh-based approaches, as a single pixel may contain both boundary and interior regions. This ambiguity complicates loss function implementation. It requires decision-making about whether a given pixel contributes to boundary constraints or PDE residuals. This issue is also present in the case of our inclined pillars, where separating between the pillars and surrounding air can be challenging due to numerical artifacts present when deciding which pixel belongs to which medium. PointNets (\hyperref[fig:pointnetc_schematic]{Fig.~\ref{fig:pointnetc_schematic}}), on the other hand, eliminate this issue, as they naturally distinguish between boundary and interior points.

Another limitation of mesh-based approaches is their poor performance when processing sparse data. They compute across the entire grid, wasting resources on empty or irrelevant regions. This inefficiency makes them less suitable for applications with highly sparse or irregularly distributed data.

\subsection{Physics-Informed Machine Learning}
\label{sec:lossFunction}

Unlike traditional supervised learning that relies solely on input-output pairs, PIML requires a specialized loss function that enforces compliance with the governing PDEs, boundary conditions, and any available reference data throughout the computational domain. This physics-constrained optimization ensures that the learned solution is physically meaningful and generalizable beyond the training data. 

The total PIML loss function is composed as a weighted sum of these residuals over their respective domains:
\begin{equation}
    \mathcal{L}(\theta) = \lambda_{i} \mathcal{L}_i + \lambda_b \mathcal{L}_b + \lambda_r \mathcal{L}_r,
\end{equation}
where $\theta$ are the trainable parameters of the network. Each term corresponds to a specific loss---$\mathcal{L}_i$ for interior points, $\mathcal{L}_b$ for boundary points, and $\mathcal{L}_r$ for reference points. The weighting factors \( \lambda_{i}, \lambda_b,\) and \( \lambda_r \) balance the contributions of the PDE, boundary, and reference losses, respectively. By using the $L^2$-norms of the residuals for the computations of the loss terms, the optimization problem is given by:
\begin{equation}
\begin{aligned}
        \theta^* = \arg \min_{\theta} \mathcal{L}(\theta) &=  \frac{1}{|\Omega|} \int_{\Omega} \left| \mathcal{R}_i(\mathbf{x}^i, \varepsilon (\mathbf{x}^i); \theta) \right|^2 \, d\mathbf{x}^i\\
        &+ \frac{1}{|\partial \Omega|} \int_{\partial \Omega} \left| \mathcal{R}_b(\mathbf{x}^b, \varepsilon (\mathbf{x}^b); \theta) \right|^2 \, d\mathbf{x}^b\\
        &+ \frac{1}{N_r} \sum_{j=1}^{N_r} \left| \mathcal{R}_r(\mathbf{x}_j^r, \varepsilon (\mathbf{x}_j^r); \hat{\mathbf{E}} (\mathbf{x}_j^r), \theta) \right|^2,
\end{aligned}
\end{equation}
where $\{\mathcal{R}_q\}_{q\in\{i,b,r\}}$ are the residual functions corresponding to the interior, boundary, and residual terms. Since exact integration is often computationally infeasible, we approximate the integral terms using numerical quadrature or Monte Carlo sampling over a discrete set of collocation points \( \{\mathbf{x}_j^i\}_{j=1}^{N_i} \subset \Omega \) for the PDE residual, and \( \{\mathbf{x}_j^b\}_{j=1}^{N_b} \subset \partial \Omega \) for the boundary residual. Similarly, we have a set of points \( \{\mathbf{x}_j^r\}_{j=1}^{N_r} \subset \Omega \) at which the reference solutions are enforced. Thus, the total number of points in each point cloud is $N_t=N_i+N_b+N_r$. A representative 2D training point cloud with $N_i=1696$, $N_b=128$, and $N_r=224$, for a total of $N_t=2048$ can be seen in \hyperref[fig:example_domain]{Fig.~\ref{fig:example_domain}a}.

The approximation of \(\mathbf{E}\) by the NN depends on the spatial coordinates \(\mathbf{x}\), the permittivity distribution \(\varepsilon(\mathbf{x})\), and the set of trainable network parameters \(\theta\). We denote this approximation as \(\mathbf{E}(\mathbf{x}; \theta)\), thereby expanding our previous definition of $\mathbf{x}$ to also include the permittivity at the given coordinate. The resulting discrete optimization problem is then given by:
\begin{equation}
    \begin{aligned}
    \label{eq:mse_training}
        \theta^* = \arg \min_{\theta}  \mathcal{L}(\theta)  & \approx \frac{\lambda_i}{N_t} \sum_{j=1}^{N_i} \left| \mathcal{R}_i(\mathbf{x}^i_j, \varepsilon (\mathbf{x}_j^i); \theta) \right|^2 \\
    &+ \frac{\lambda_b}{N_b} \sum_{j=1}^{N_b} \left| \mathcal{R}_b(\mathbf{x}^b_j, \varepsilon (\mathbf{x}^b_j); \theta) \right|^2 \\
    &+ \frac{\lambda_r}{N_r} \sum_{j=1}^{N_r} \left| \mathcal{R}_r(\mathbf{x}^r_j; \hat{\mathbf{E}}(\mathbf{x}_j^r), \theta) \right|^2
    \end{aligned}
\end{equation}

where, for the 2D case, we have

\begin{equation}
    \begin{aligned}
        &\mathcal{R}_i(\mathbf{x}^i_j, \varepsilon (\mathbf{x}^i_j); \theta) = \nabla^2 \mathbf{E}(\mathbf{x}^i_j; \theta)  +  \omega^2 \varepsilon(\textbf{x}^i_j)\mathbf{E}(\mathbf{x}^i_j, \theta) \\
        &\mathcal{R}_b(\mathbf{x}^b_j; \theta) = \sum_{\alpha=0}^1\partial^{\alpha} \mathbf{E}(x_j^{\Gamma_a}, z_j^b; \theta) - \partial^{\alpha} \mathbf{E}(x_j^{\Gamma_b}, z_j^b; \theta) \\
        &\mathcal{R}_r(\mathbf{x}^r_j; \hat{\mathbf{E}}(\mathbf{x}^r_j), \theta) =  \mathbf{E}(\mathbf{x}^r_j; \theta) - \hat{\mathbf{E}}(\mathbf{x}^r_j). \\
    \end{aligned}
\end{equation}

Here, $\hat{\mathbf{E}}(\mathbf{x}_j^r)$ is the reference solution at the given point $\mathbf{x}_j$. For the 3D case, $\mathcal{R}_i(\mathbf{x}^i_j, \varepsilon (\mathbf{x}^i_j); \theta)$ is substituted by the relevant formulation given by \eqref{eq:helmholtz} .

\subsection{Implementation}
\label{sec:Implementation}

Before training, a single point cloud is sampled for each domain in \(\mathcal{D}_t\), representing a unique material distribution. The point clouds were sampled using Halton sequences, with reference point locations fixed and shared across all domains in \(\mathcal{D}_t\). In contrast, during testing, each domain in \(\mathcal{D}_v\) is subdivided into multiple point clouds. This leads to smoother, more consistent validation errors and images. 

The Mean Absolute Percentage Error ($\mathrm{MAPE}$) of the amplitude of the predicted electric field was used as a near-field validation metric. When looking at pointwise differences of the near-field, we use the Absolute Error ($\mathrm{AE}$) and Mean Absolute Error ($\mathrm{MAE}$) to maintain consistency with the visualizations of the near-field errors. The Absolute Percentage Error ($\mathrm{APE}$) was employed when finding errors between different diffraction efficiencies of transmitted light. Since the period of the metasurface is smaller than the exposure wavelength, there are only one or a few propagating diffraction orders in the transmitted light. Subsequently, we analyze the diffraction efficiency of $0^{th}$ diffraction order by near-to-far-field transformation using the fourier transform.

Unless otherwise stated, each PIPN is trained using the Adam optimizer~\cite{kingma_adam_2017} for 50,000 iterations. Testing errors were computed at the end of each training. A batch size of 8 was used. A polynomial learning rate decay was employed, starting from an initial learning rate of \(10^{-3}\) and decaying to a final learning rate of \(8 \times 10^{-5}\), with a polynomial decay coefficient of 4.5. A \texttt{sine} activation function was used in all layers except the final one. The weighting factors specified in \hyperref[sec:lossFunction]{Sec.~\ref*{sec:lossFunction}} were set to 1, 1, and 10 for $\lambda_i, \lambda_b$, and $\lambda_r$ respectively, as consistent with prior studies~\cite{kashefi_prediction_2023}. The implementation was done in Python using TensorFlow 1.x, and all training and inference was conducted on an NVIDIA Quadro P5000 GPU with 16 GB of VRAM.

\section{Results}
\label{sec:Results}

To establish the optimal network configuration used in subsequent analyses, several key architectural and preprocessing considerations were investigated. First, we examined variations to the network structure by replacing the 1D convolutional layers after aggregation with dense layers (PointNetMLP) or a Chebychev-KAN (PointNetCheby), see \hyperref[app:varying_pointnet_architectures]{App.~\ref{app:varying_pointnet_architectures}}. Here we find that a PointNetMLP with $N_\theta=252 \times 10^3$ trainable parameters yields a good balance between computational cost and testing error for the 2D case. Furthermore, data normalization is an important pre-processing step in traditional deep learning. This typically involves scaling the input features of a dataset so that they have similar magnitudes and ranges when compared to the output variables \cite{glorot_understanding_2010}. The specific input scaling constant necessary for achieving converging results in our experiments are thoroughly discussed in \hyperref[app:Dependance_on_Input_Scale]{App.~\ref{app:Dependance_on_Input_Scale}}. Here we find that a scaling factor of approximately 21 is necessary for our problem to give plausible approximations. Lastly, the selection of input point clouds is another critical factor for the performance and computational efficiency of PIPNs. Our investigation (presented in \hyperref[app:dependance_on_sensor_points_and_material_interface]{App.~\ref{app:dependance_on_sensor_points_and_material_interface}} \hyperref[fig:raytune_n_sep_generalized]{Fig.~\ref{fig:raytune_n_sep_generalized}}) reveals that optimal performance is achieved with $N_i=1696$, $N_b=128$, and $N_r=224$, which yields the lowest near-field $\mathrm{MAPE}$.

\begin{table}[!ht]
\centering
\caption{Summary of results presented in \hyperref[sec:Results]{Sec.~\ref{sec:Results}}. Error metrics represent the mean and standard deviation computed over the testing dataset $\mathcal{D}_v$ for both near-field predictions and diffraction efficiency calculations.}
\label{tab:summary}
\resizebox{3.5in}{!}{%
\renewcommand{\arraystretch}{1.2}
\begin{tabular}{lccc}
\toprule
& \multicolumn{2}{c}{\textbf{SiO\(_2\)}} & \textbf{TiO\(_2\)} \\
\cmidrule(lr){2-3}\cmidrule(lr){4-4}
\textbf{Dataset} 
  & \hyperref[sec:low_refractive_index]{Case I} 2D
  & \hyperref[sec:high_refractive_index]{Case II} 3D
  & \hyperref[sec:3d_homogeneous]{Case III} 2D\\
\midrule
\textbf{Trainable Network Parameters [$N_{\theta}$]} 
  & 252,248
  & 448,846 
  & 447,594 \\

\textbf{Training Time [$hours$]}
  & \(1.96\) 
  & \(24.6\)
  & \(3.92\) \\

$\textbf{Near-Field MAPE}$ [$\%$]
  & \(1.69 \pm 0.43\)
  & \(3.84 \pm 0.13\) 
  & \(10.72 \pm 5.64\)\\

\textbf{Diffraction Efficiency APE [$\%$]}
  & \(1.95 \pm 1.10\)
  & \(3.91 \pm 4.18\) 
  & \(6.60 \pm 3.37\) \\
\bottomrule
\end{tabular}
}
\end{table}

The hyperparameters used in the following subsections are informed by this analysis; \hyperref[tab:summary]{Tab.~\ref*{tab:summary}} summarizes the corresponding results.

\subsection*{Case I - 2D SiO\texorpdfstring{\textsubscript{2}}{2} Metasurface}
\label{sec:low_refractive_index}

\begin{figure*}
    \centering
    \includegraphics[width=7in]{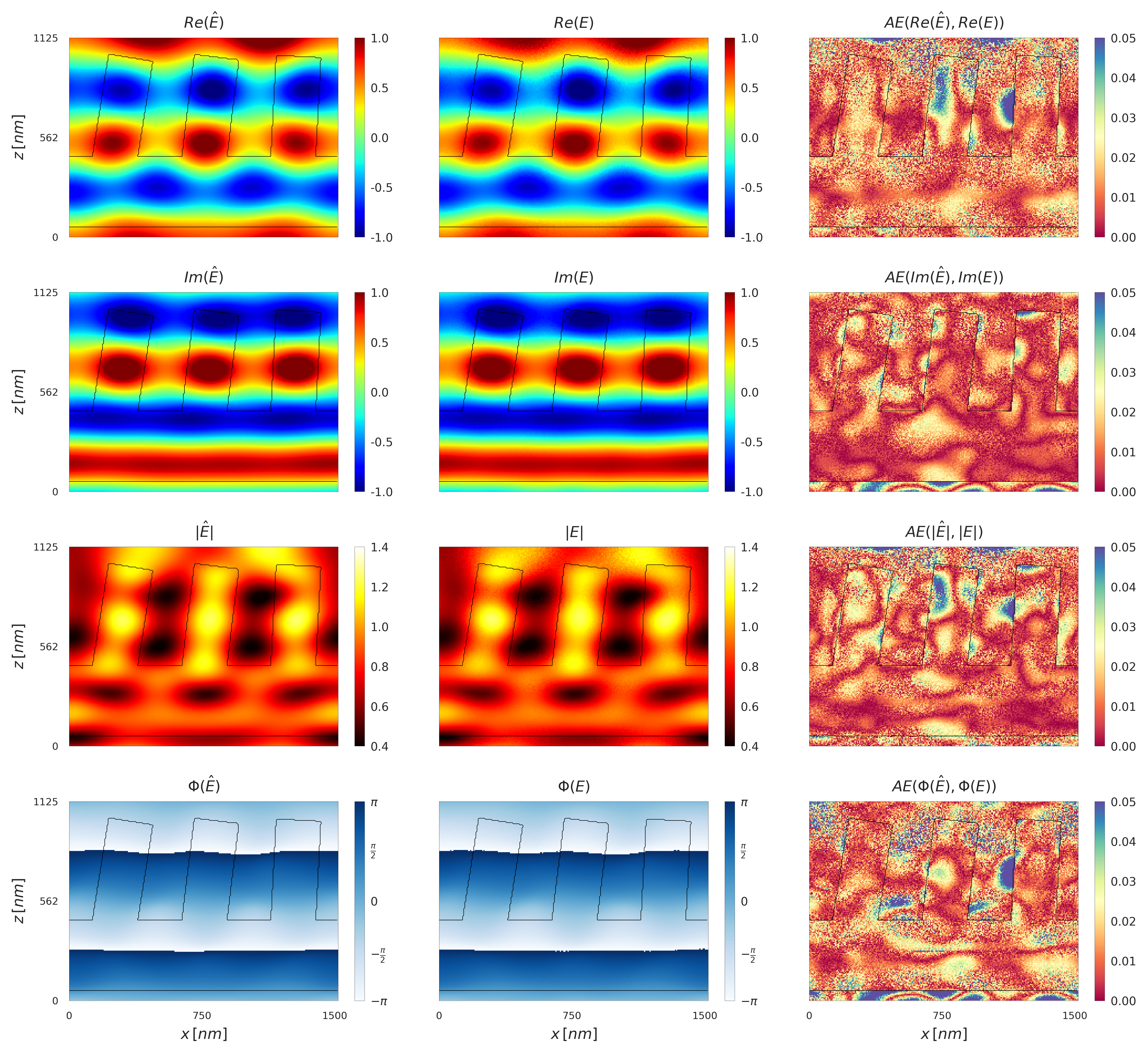} 
    \caption{A representative approximate solution (near-field $\mathrm{MAPE}=1.59\%$), the reference solution, and its errors of a representative test domain of the $SiO_2$ dataset. The first column shows the real, imaginary, amplitude, and phase of the reference electric field. The second column shows the PIPN solution, and the third column shows the $\mathrm{MAE}$ between them. The metasurface contour is highlighted in black.}
    \label{fig:exampleSolution}
\end{figure*}

\hyperref[fig:exampleSolution]{Fig.~\ref{fig:exampleSolution}} depicts the near-field of a representative test domain. Columns show reference, predicted electric fields and $\mathrm{AE}$ between the two; rows show the real, imaginary, amplitude and phase.

The near-field $\mathrm{MAPE}$ for the test domain in \hyperref[fig:exampleSolution]{Fig.~\ref{fig:exampleSolution}} is $1.59\%$, whereas the near-field $\mathrm{MAPE}$ for $ \mathcal{D}_v $ is $1.69\pm0.43\%$. Similarly the diffraction efficiency $\mathrm{APE}$ for all test domains is $1.95\pm1.1\%$. The training took around 2 hours and inference for an entire testing domain is on the scale of $100ms$. The $\mathrm{MAE}$ for $\mathcal{D}_v$ is around $0.02$ with regions around the material interface and boundaries reaching values as high as $0.08$. A detailed plot showing the dependence of the $\mathrm{MAE}$ with respect to the distance from the nearest material interface can be seen in \hyperref[app:dependance_on_sensor_points_and_material_interface]{App.~\ref{app:dependance_on_sensor_points_and_material_interface}} in \hyperref[fig:distances_to_material_interface_and_reference_point_sio2]{Fig. \ref{fig:distances_to_material_interface_and_reference_point_sio2}}. The results demonstrate that the pointwise $\mathrm{AE}$ decreases linearly with increasing distance to the nearest material interface, reducing by approximately 50\% from 0.028 to 0.015 as the distance approaches 200~$\text{nm}$. Conversely, the pointwise $\mathrm{AE}$ exhibits an exponential increase with distance to the nearest training reference point, more than doubling from approximately 0.018 to 0.042 as the distance reaches 130~$\text{nm}$ (see \hyperref[fig:distances_to_material_interface_and_reference_point_sio2]{Fig. \ref{fig:distances_to_material_interface_and_reference_point_sio2}}).

\subsection*{Case II - 3D SiO\texorpdfstring{\textsubscript{2}}{2} Metasurface}
\label{sec:3d_homogeneous}

\begin{figure}
    \centering
    \includegraphics[width=3.5in]{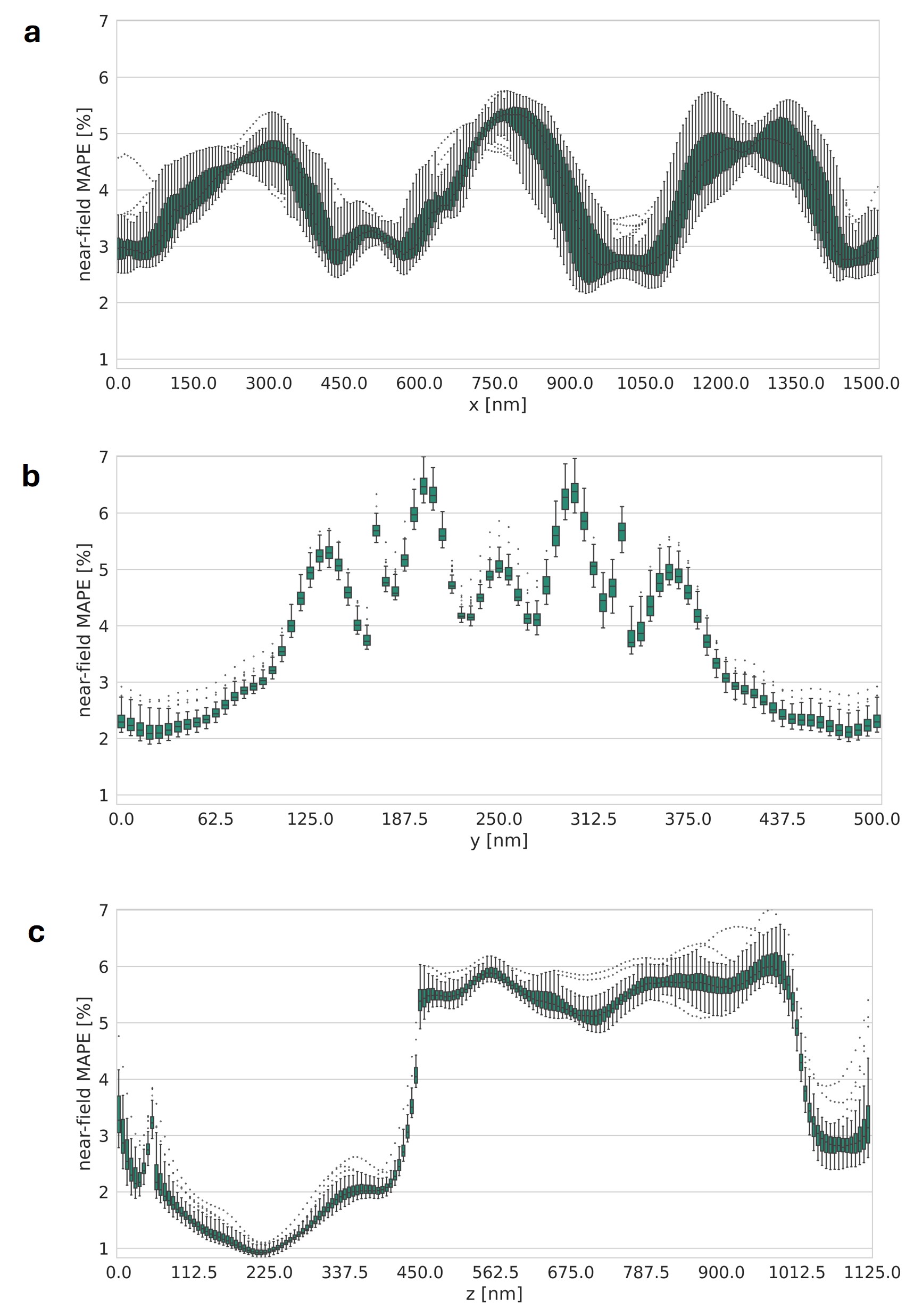} 
    \caption{The near-field $\mathrm{MAPE}$ for the 3D test dataset when slicing in the (a) $x$-dimension (i.e.\ looking at the results in the $yz$-plane), (b) $y$-dimension or (c) $z$-dimension.  Metasurface geometry and material as indicated in \hyperref[fig:example_domain]{Fig.~\ref{fig:example_domain}b}.}
    \label{fig:3d_case}
\end{figure}

The network output now includes the $E_x$, $E_y$, and $E_z$ fields (instead of just $E_y$ for the 2D case) according to \eqref{eq:helmholtz}. The training iterations are increased to $100,000$, and the point cloud is doubled in size (i.e $N_i=3392$, $N_b=256$ and $N_r=448$) to account for the larger domain. Furthermore we increase the network size to $N_\theta = 448\times10^3$ traianble parameters. Due to this, the memory requirements went up significantly and the batch size had to be decreased to 2 for us to still be able to train on a 16GB VRAM GPU.\ The training time increased by an order of magnitude to approximately 24 hours.
\begin{figure*}
    \centering
    \includegraphics[width=7in]{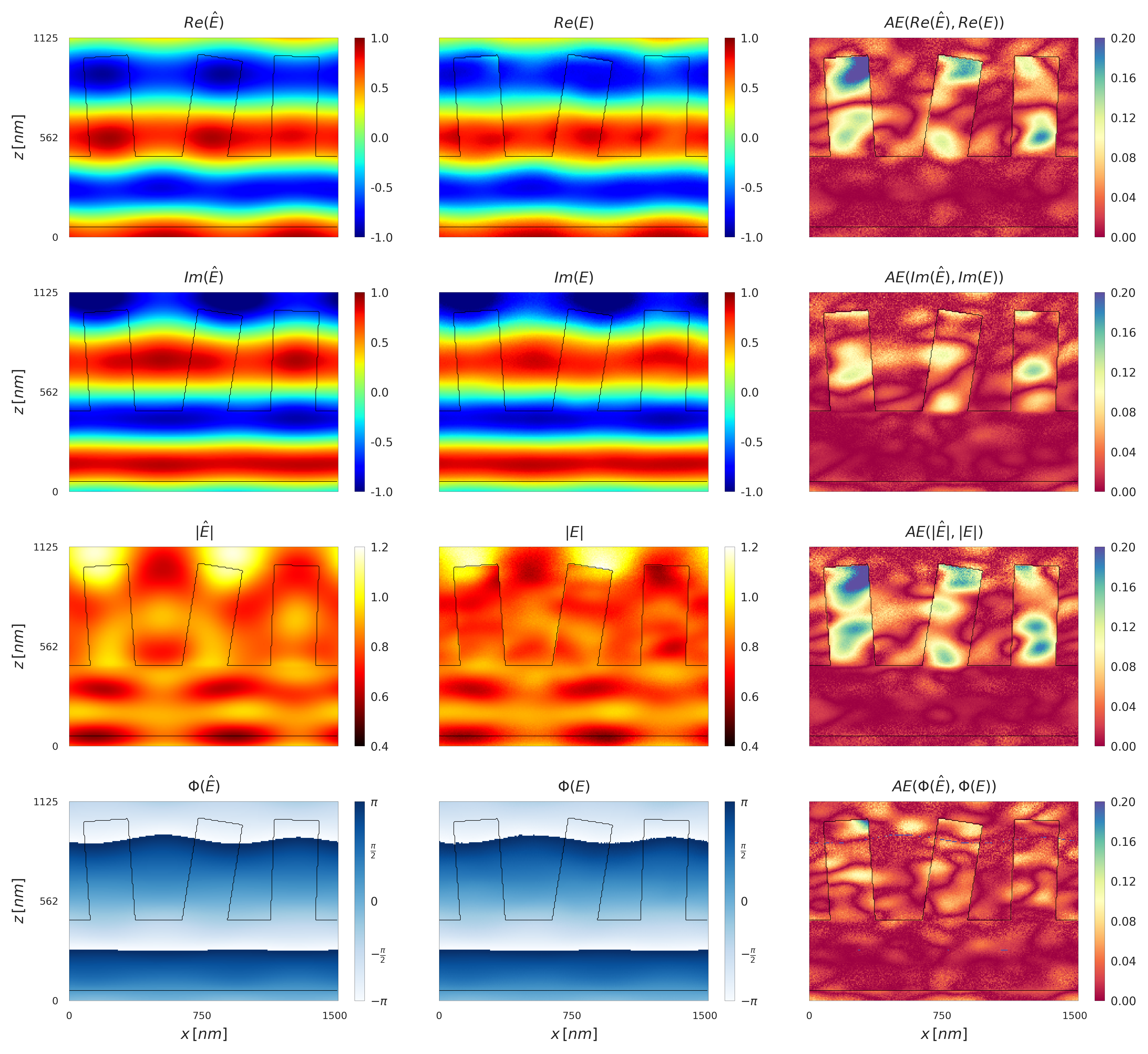} 
    \caption{Representative solution for the 3D $\mathrm{SiO_2}$ $E_y$ field at $y\approx250 \, \text{nm}$, with near-field $\mathrm{MAPE}=4.04\%$. Metasurface geometry and material as indicated in \hyperref[fig:example_domain]{Fig.~\ref{fig:example_domain}b}.}
    \label{fig:exampleSolution3Dsl39}
\end{figure*}
\hyperref[fig:exampleSolution3Dsl39]{Fig.~\ref{fig:exampleSolution3Dsl39}} illustrates the reference solution and predicted near-fields of a representative test domain slice ($y\approx250 \, \text{nm})$. The entire domain has a near-field $\mathrm{MAPE}$ of $4.04\%$, whereas  set $ \mathcal{D}_v $ has a near-field $\mathrm{MAPE}$ of $3.04\pm0.09\%$, and a diffraction efficiency $\mathrm{APE}$ of $1.12\pm0.28\%$. 
\begin{figure*}
    \centering
    \includegraphics[width=7in]{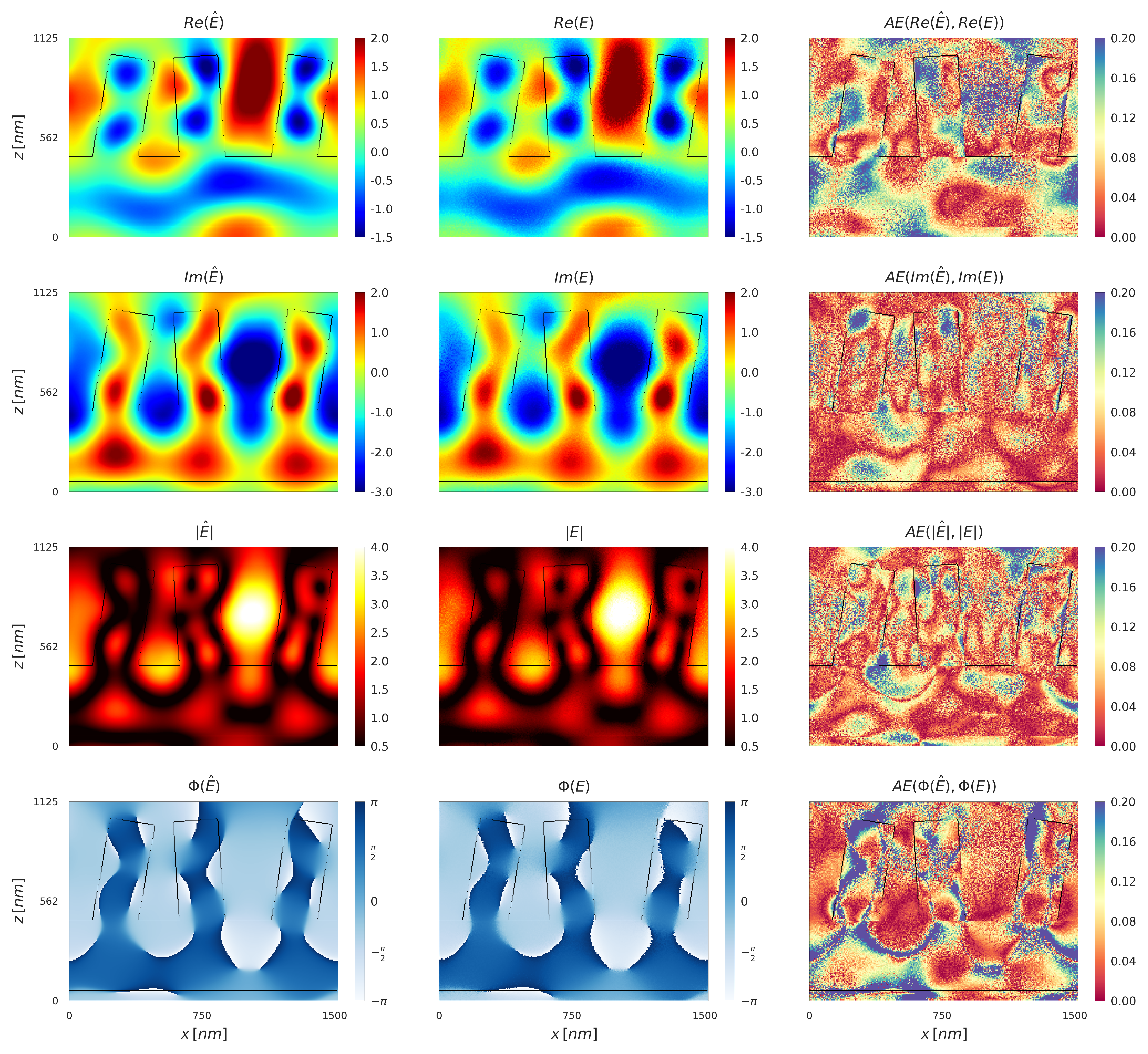} 
    \caption{Representative solution for the 2D $\mathrm{TiO_2}$ dataset, with near-field $\mathrm{MAPE} =10.01\%$.}
    \label{fig:exampleSolutionLargerRefractive}
\end{figure*}
The near-field $\mathrm{MAPE}$ distribution over all slices in the $x$-, $y$-, and $z$-directions can be seen in \hyperref[fig:3d_case]{Fig.~\ref{fig:3d_case}}. 
The increased error, starting at the material interfaces and going into the pillars, is observable in \hyperref[fig:3d_case]{Fig.~\ref{fig:3d_case}a}. No error jump on the material interface is discernible, due to the averaging over $\mathcal{D}_v$ (i.e., the angles $\varphi_0, \varphi_1, \varphi_2$ in the $x$-direction of the pillars are different, and thus no clear transition can be seen). The error profile is symmetric, reflecting the symmetry of the material properties with respect to the pillar center in the y-direction (\hyperref[fig:3d_case]{Fig.~\ref{fig:3d_case}b}). A steep decrease in near-field $\mathrm{MAPE}$ below approximately 125~$\text{nm}$ and above 375~$\text{nm}$ coincides with the substrate/pillar interface.

\subsection*{Case III - 2D TiO\texorpdfstring{\textsubscript{2}}{2}-on-SiO\texorpdfstring{\textsubscript{2}}{2} Metasurface}
\label{sec:high_refractive_index}

We use a network of size $N_\theta = 447 \times 10^3$. Besides this, no network parameters were changed compared to Case I. The training took around 4 hours. The near-field approximation of a representative test domain with $\mathrm{MAPE}=10.01\%$ is shown in \hyperref[fig:exampleSolutionLargerRefractive]{Fig.~\ref{fig:exampleSolutionLargerRefractive}}. The testing domains $ \mathcal{D}_v $ have a near-field $\mathrm{MAPE}$ of $10.72 \pm 5.64\%$ and a diffraction efficiency $\mathrm{APE}$ of $6.6 \pm3.37 \%$. The largest point wise $\mathrm{AE}$ is found in the regions between the pillars and around the material interface. Moreover, the predicted solution appears grainy compared to \hyperref[fig:exampleSolution]{Fig.~\ref{fig:exampleSolution}}, suggesting that this artifact is exaggerated in the presence of larger material contrast. The $\mathrm{MAE}$ for the testing dataset is around $0.041$ with select regions around the material interface and boundaries reaching values as high as $0.2$.

\section{Discussion}
\label{sec:discussion}

The results presented in \hyperref[sec:Results]{Sec.~\ref{sec:Results}} demonstrate the robust generalization capabilities of the PIPN framework for metasurfaces with varying refractive indices. This establishes that variable PDE parameters can be effectively encoded within the PIPN architecture. Future investigations should examine the performance characteristics of PIPNs that incorporate spatially and temporally dependent PDE parameters across a wide variety of PDE problems. Additionally, a particular area of future study for PIPN applications should be the generalization across diverse domain geometries, as opposed to the square domains exclusively considered in this study. Across all cases, a clear correlation between material contrast and prediction accuracy is evident: the SiO\(_2\) (Case I/II) dataset achieved the best performance, while the higher contrast TiO\(_2\) (Case III) system exhibited significantly increased errors. This is to be expected because NNs typically perform better when approximating smooth functions~\cite{pmlr-v97-rahaman19a}. We note that the largest error is observed near and within the meta-atom layer, where local field enhancement and strong scattering effects cause significant field perturbations (see \hyperref[fig:3d_case]{Fig.~\ref{fig:3d_case}c}). Therefore, the concentration of points around the meta-atom layer is not guaranteed to be large enough to successfully capture the interface. Future work should investigate adaptive point sampling strategies~\cite{wu_comprehensive_2023}, which have shown promise in mitigating this issue. 

The 2D problems (Case I/III) have a small training time of only a few hours. Here, the computational requirements were low enough in both cases that with a decrease in batch size and training point cloud size, the networks could have been trained on 8GB VRAM GPUs without a significant increase in testing error. A steep increase in training time was observed for Case II\; training took around 24 hours due to the increased computational requirements, which stemmed from the vectorial formulation of the Helmholtz equation~\eqref{eq:helmholtz}. Nonetheless, our NN shows great promise in terms of the required computational resources compared to state-of-the-art mesh-based PINN approaches~\cite{medvedev_3d_2024_1}. While testing errors between the U-Net~\cite{medvedev_3d_2024_1} and this PIPN are comparable, the PIPN demonstrates significant computational advantages. Our approach can be trained on a NVIDIA Quadro P5000 GPU with 16 GB of VRAM in approximately 24 hours, whereas the mesh-based counterpart requires a NVIDIA A80 with 80GB VRAM and training times of 90 hours for high-resolution images. For context, the mesh-based approach uses a resolution of $\text{pixel}^3 = 8\text{nm}^3$, which is comparable to our effective resolution of $\text{pixel}^3=6.25\text{nm}^3$.

Additionally, for all three cases, an increase in error can be seen at the edges of the domain. This can be explained by the fact that the point density is smallest at the edge. Future work---similar to \hyperref[app:varying_pointnet_architectures]{App.~\ref{app:varying_pointnet_architectures}}---should be done on adapting the PointNet architecture to further mitigate computational requirements; also, to test whether or not the proposed architectural changes yield improved performances for other PDE problems. We have further tested architectures involving Fourier-KAN~\cite{imran_fourierkan_2024}, RBF-KAN \cite{li_kolmogorov-arnold_2024}, and B-Spline-KAN~\cite{liu_kan_2024}, as well as one implementation that replaces the convolutional kernels used in the PIPN framework by B-Splines. No converging results were obtained, and thus were omitted from this work. However, when used as a standalone network to solve the ungeneralized problem, Fourier-KAN showed great promise with lower training time, network size, and testing errors. 

Lastly, further investigation is required to characterize the failure modes of PIPNs. As demonstrated in \hyperref[app:Dependance_on_Input_Scale]{App.~\ref{app:Dependance_on_Input_Scale}}, the convergence of PIPN to a physically meaningful approximation exhibits strong sensitivity to the input coordinate scaling. Consequently, conventional normalization strategies that scale NN inputs to magnitudes comparable to the outputs~\cite{cuomo_scientific_2022, wang_experts_2023, glorot_understanding_2010} may not be universally applicable across all PIML formulations.

\section{Conclusion}
\label{sec:Conclusion}

This work examined the potential of PIPNs in solving the optical scattering problem involving all-dielectric metasurfaces that contain nanopillars with varying inclinations and refractive indices. The proposed method is a mesh-free, PIML approach that achieves moderate accuracy and generalizability while relying only on weak supervision from sparse reference solution while requiring a fraction of the computational resources of its mesh-based counterparts~\cite{medvedev_physics-informed_2025}. In particular, our PIPN enables us to encode the spatially varying PDE parameter, which allows the network to differentiate domains based on the underlying material distribution. Furthermore, although not an intrinsic requirement, the inclusion of reference points may prove fruitful for optical applications where the source is not fully defined, but experimentally-collected reference data is available. While the framework exhibited promising performance in various scenarios, it encountered notable limitations, especially in regards to domains characterized by large material contrast. These limitations produced noticeable artifacts and increased the prediction errors, both of which highlight the challenges that PIPNs face when approximating sharp transitions. Hyperparameter tuning provided valuable insights into optimal input point cloud configuration and PointNet architectures. Although further hyperparameter optimization was outside the scope of this study, the limited improvements observed suggest that the current parameters already achieve near-optimal performance.

\appendix
\label{sec:appendix}

\subsection{Error Dependence on the Input Scale}
\label{app:Dependance_on_Input_Scale}

\begin{figure}[!ht]
    \centering
    \includegraphics[width=3.5in]{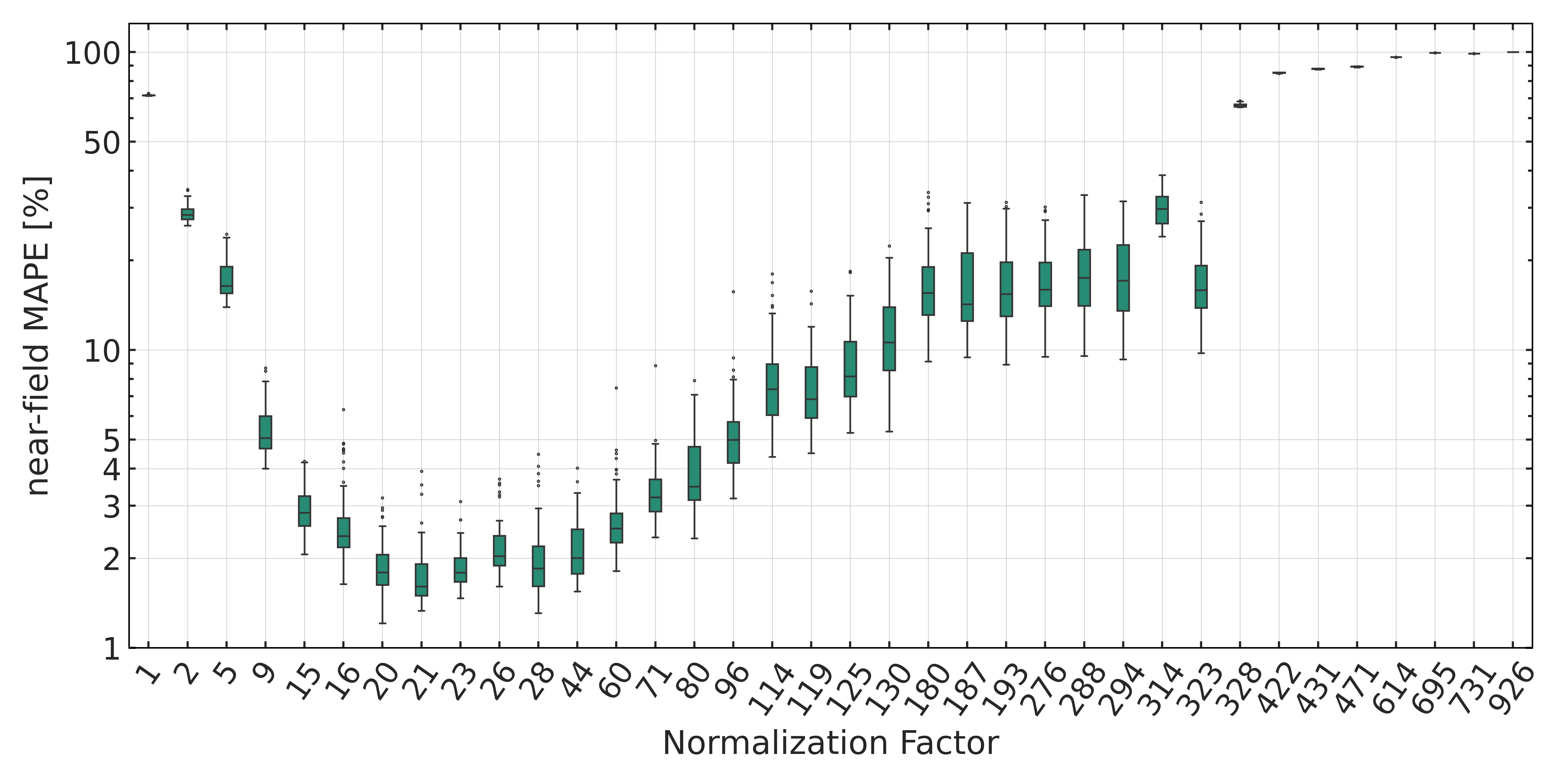}
    \caption{Relationship between the normalization factor (of the input spatial coordinates and wavelength) and the near-field $\mathrm{MAPE}$ of the testing domains.}
    \label{fig:raytune_norm_factor}
\end{figure}

\hyperref[fig:raytune_norm_factor]{Fig.~\ref{fig:raytune_norm_factor}} shows the near-field $\mathrm{MAPE}$ distribution over all testing domains when the input coordinates and wavelength are scaled by different constants.
The best results are achieved with a scaling factor around 21 (note the non-uniform $x$-axis). Using this, the input $x$-coordinates get scaled to around [-35, 35], and the $z$-coordinates are scaled to around [-18, 18]. This is still an order of magnitude larger than the output of the network (real and imaginary parts of the electric field); that output is on a scale of around [-1.1, 1.1] for Case I/II, and within the range [-3.7, 2.3] for Case III dataset. The $x$- and $z$-coordinates are an order of magnitude larger than the permittivity, which is in a range of [1.0, 2.1] for Case I/II and [1.0, 6.4] for Case III.

\subsection{Comparison of Assorted PointNet Architectures}
\label{app:varying_pointnet_architectures}

\begin{figure}[!ht]
    \centering
    \includegraphics[width=3.5in]{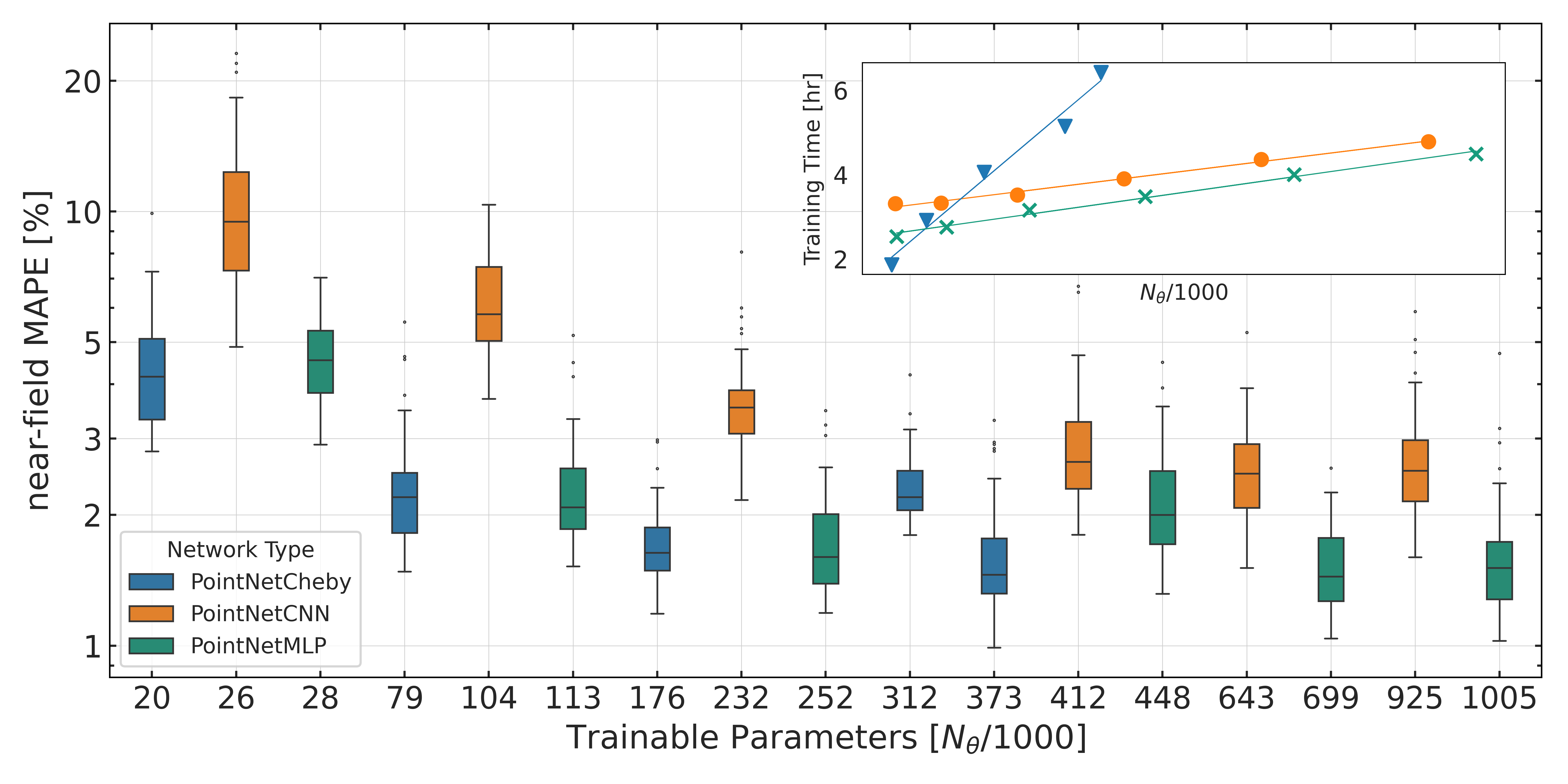}
    \caption{Box plots of the test near-field $\mathrm{MAPE}$ versus the network size ($N_{\theta}$) for three different variations of the PointNet architecture. The inset plot shows the relative training time (hours) versus the same network sizes.}
    \label{fig:raytune_different_pointnets}
\end{figure}

\hyperref[fig:raytune_different_pointnets]{Fig.~\ref{fig:raytune_different_pointnets}} compares the performance of three PointNet architecture variations across different network sizes, where each network size is scaled by proportionally decreasing the number of nodes in each layer. PointNetCNN (orange) represents the unchanged architecture depicted in \hyperref[fig:pointnetc_schematic]{Fig.~\ref{fig:pointnetc_schematic}}; PointNetMLP (green) replaces the second half of convolutional kernels (as indicated by the dashed squared in \hyperref[fig:pointnetc_schematic]{Fig.~\ref{fig:pointnetc_schematic}}) with a three-layer dense neural network. PointNetCheby (blue) similarly substitutes the second half with a three-layer KAN that employs Chebyshev basis functions.

All three architectures demonstrate continuous improvement in mean near-field MAPE with increasing network size, though they exhibit distinct performance characteristics. PointNetCheby achieves the best accuracy across most network sizes, reaching a minimum mean near-field MAPE of $1.71 \pm 0.37 \%$ at approximately 373$\times 10^3$ parameters before GPU VRAM limitations are encountered. PointNetMLP shows comparable performance to PointNetCheby at smaller network sizes, achieving $1.66 \pm 0.53\%$ mean near-field MAPE at 252$\times10^3$ parameters. In contrast, PointNetCNN consistently exhibits the highest error rates, with its best performance of $2.69 \pm 0.94\%$ mean near-field MAPE occurring at the largest tested network size: 995$\times10^3$ parameters.

The inset plot reveals linear scaling of training time with network size across all architectures. The training times range from 3 to 6.5 hours, with the PointNetCheby demonstrating the worst scaling behavior and the longest training times. PointNetMLP and PointNetCNN exhibit similar scaling behavior to each other, with PointNetMLP consistently requiring about 0.5 hours less training time and reaching maximum training times of approximately 4.5 hours.

\subsection{Error Dependence on Reference Points and Distance to Material Interface}
\label{app:dependance_on_sensor_points_and_material_interface}

When working with point clouds, the selection of points can significantly impact performance. The three key components within our point cloud are: boundary points, interior points, and reference points. Intuitively, the optimal configuration would be achieved by maximizing the number of points in each category, thus giving the NN the largest possible amount of information about the system. However, increasing the number of reference points raises dependence on a priori data which can be costly to obtain. \\

\begin{figure}[!ht]
    \centering
    \includegraphics[width=3.5in]{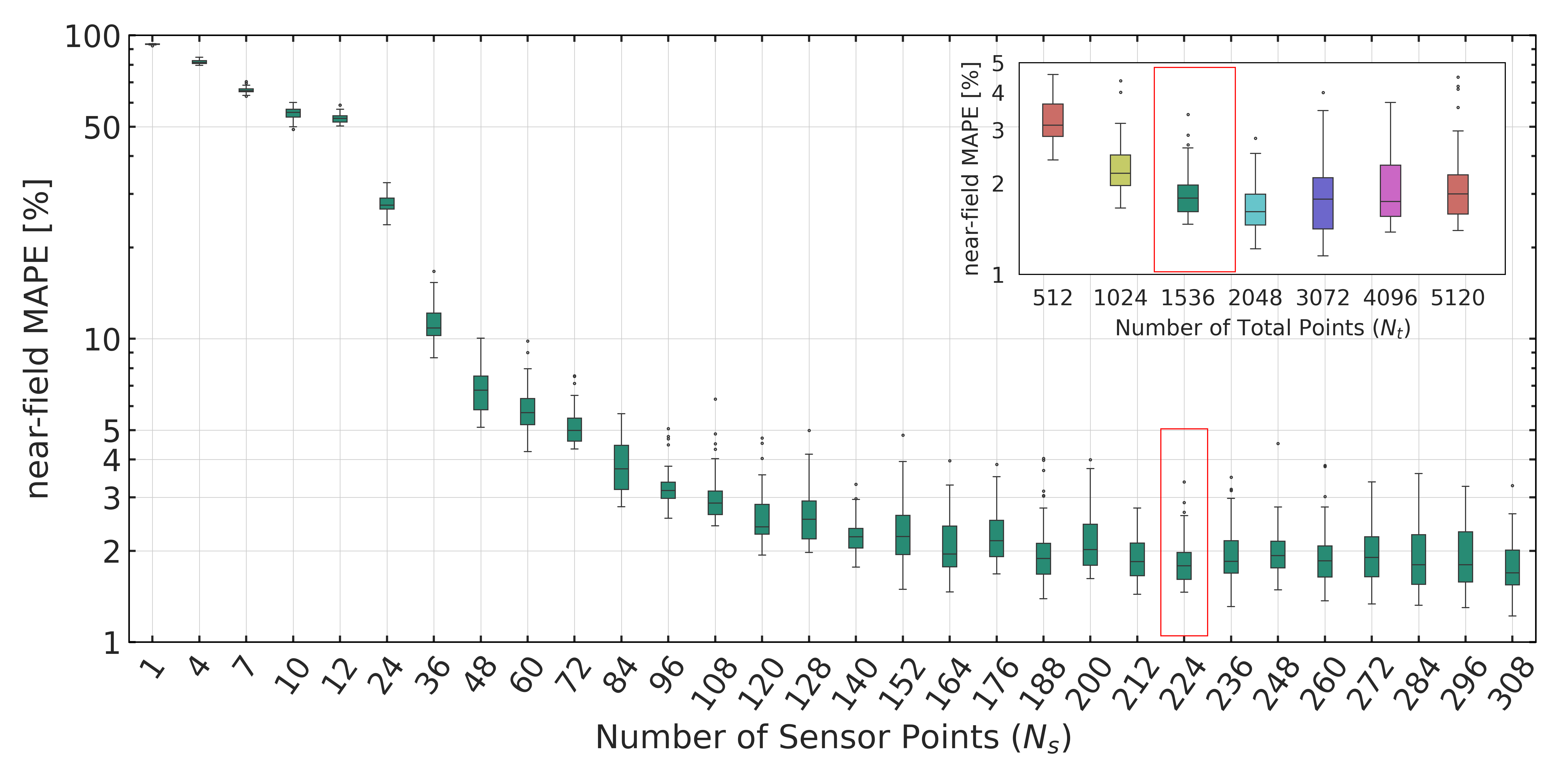}
    \caption{Box plots of the test near-field $\mathrm{MAPE}$ versus the number of reference points ($N_r$) used during training. The inset plot shows how varying the number of total points ($N_t$) used for training affects the near-field $\mathrm{MAPE}$ at a constant $N_r=224$.}
    \label{fig:raytune_n_sep_generalized}
\end{figure}

\hyperref[fig:raytune_n_sep_generalized]{Fig.~\ref{fig:raytune_n_sep_generalized}} shows the dependence of the near-field $\mathrm{MAPE}$ on the number of reference points and total points used during training. The figure depicts the $\mathrm{MAPEs}$ distribution over the 50 testing domains when varying the number of reference points used during training ($N_t=1536$, $N_b=128$ and $N_i=1536-128-N_r$). Results improve until $N_r=224$, when a near-field $\mathrm{MAPE}$ of $1.88 \%$ is obtained; after that, the mean $\mathrm{MAPE}$ does not change significantly. The inset plot depicts the error distribution when varying $N_t$ ($N_r=224$, $N_b=128$ and $N_i=N_t-128-224$). The average $\mathrm{MAPE}$ decreases until $N_t=2048$, at which a near-field $\mathrm{MAPE}$ of $1.67 \%$ is obtained. After that point, it increases slightly until $N_t=5120$. This is likely due to small network size, which prevents the network from effectively leveraging additional information. However, the plateau could also be explained by the network overfitting on the reference points and neglecting physics.

\begin{figure}[!ht]
    \centering
    \includegraphics[width=3.5in]{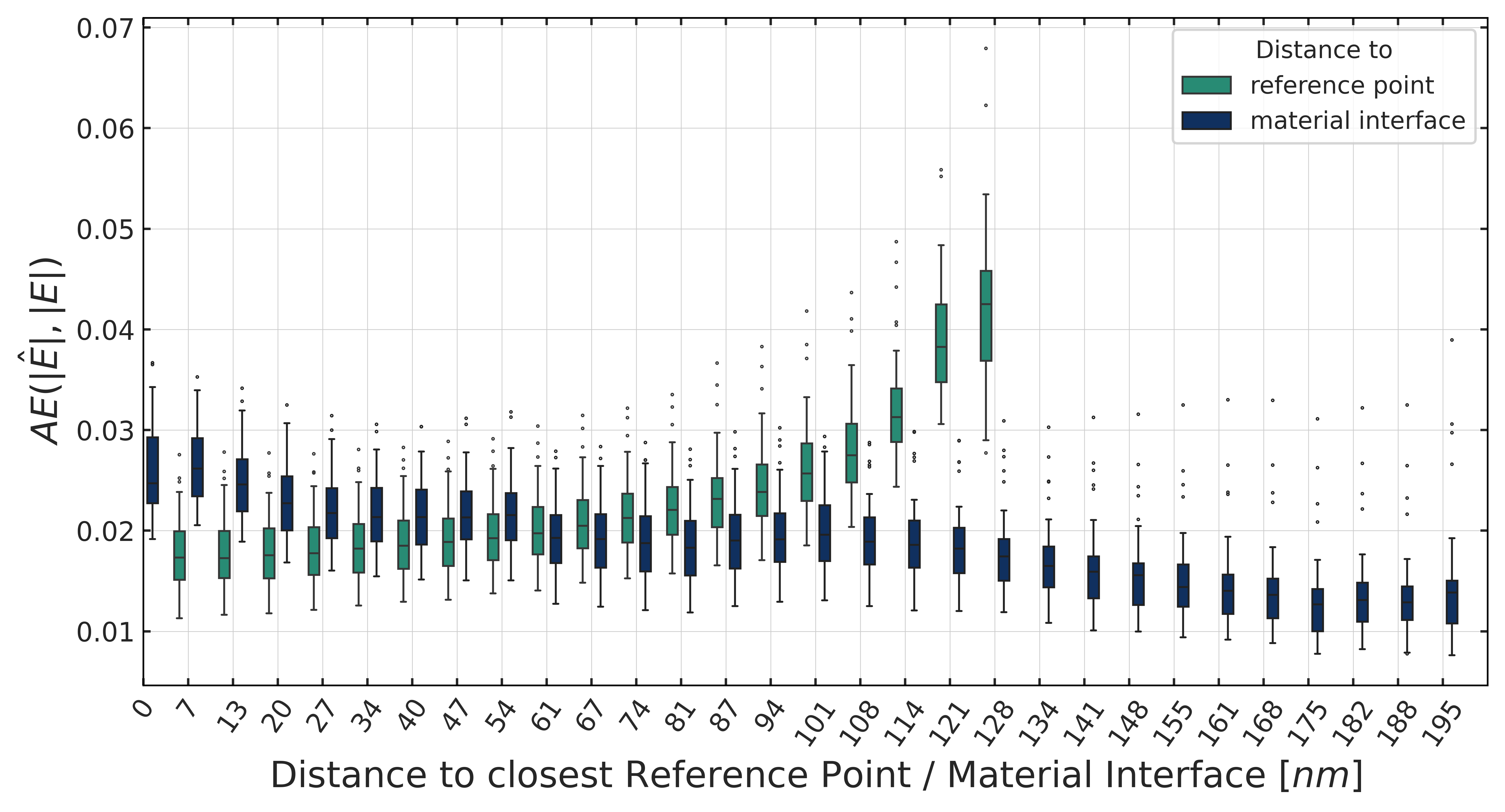}
    \caption{Correlation between the near-field $\mathrm{MAE}$ (averaged over the testing set $\mathcal{D}_v$ from Case I - 2D SiO$_2$ Metasurface) at each test domain point and its minimum distance from the corresponding training reference points (green) and material interface (blue).}
    \label{fig:distances_to_material_interface_and_reference_point_sio2}
\end{figure}

\hyperref[fig:distances_to_material_interface_and_reference_point_sio2]{Fig.~\ref{fig:distances_to_material_interface_and_reference_point_sio2}} shows the near-field $\mathrm{MAE}$ distribution at each point versus the minimum distance to the closest material interface and reference training points. The $\mathrm{MAE}$ increases continuously as the distance to the closest reference point increases. Most points are 60~$\text{nm}$ away from the nearest reference solution, with decreasing numbers for points that are either closer or further away from reference points. The near-field $\mathrm{MAE}$ distribution at each point continuously decreases as the distance to the closest material interface increases. Most points in the point cloud are within 100~$\text{nm}$ of a material interface.

\section*{Acknowledgment}
\label{sec:acknowledgment}

We thank Prof.\ Dr.\ Daniel Tenbrinck and Dr.\ Andreas Erdmann for their guidance and valuable assistance in selecting this research topic. We also thank Dr. Christopher Straub for his thoughtful reviews.
\bibliographystyle{IEEEtran}
\footnotesize
\bibliography{references}

\begin{thebibliography}{10}
\providecommand{\url}[1]{#1}
\csname url@samestyle\endcsname
\providecommand{\newblock}{\relax}
\providecommand{\bibinfo}[2]{#2}
\providecommand{\BIBentrySTDinterwordspacing}{\spaceskip=0pt\relax}
\providecommand{\BIBentryALTinterwordstretchfactor}{4}
\providecommand{\BIBentryALTinterwordspacing}{\spaceskip=\fontdimen2\font plus
\BIBentryALTinterwordstretchfactor\fontdimen3\font minus
  \fontdimen4\font\relax}
\providecommand{\BIBforeignlanguage}[2]{{%
\expandafter\ifx\csname l@#1\endcsname\relax
\typeout{** WARNING: IEEEtran.bst: No hyphenation pattern has been}%
\typeout{** loaded for the language `#1'. Using the pattern for}%
\typeout{** the default language instead.}%
\else
\language=\csname l@#1\endcsname
\fi
#2}}
\providecommand{\BIBdecl}{\relax}
\BIBdecl

\bibitem{neshev_optical_2018}
D.~Neshev and I.~Aharonovich, ``\BIBforeignlanguage{en}{Optical metasurfaces:
  new generation building blocks for multi-functional optics},''
  \emph{\BIBforeignlanguage{en}{Light: Science \& Applications}}, vol.~7,
  no.~1, p.~58, Aug. 2018, publisher: Nature Publishing Group.

\bibitem{kuznetsov2024roadmap}
A.~I. Kuznetsov, M.~L. Brongersma, J.~Yao, M.~K. Chen, U.~Levy, D.~P. Tsai,
  N.~I. Zheludev, A.~Faraon, A.~Arbabi, N.~Yu \emph{et~al.}, ``Roadmap for
  optical metasurfaces,'' \emph{ACS photonics}, vol.~11, no.~3, pp. 816--865,
  2024.

\bibitem{yu2013flat}
N.~Yu, P.~Genevet, F.~Aieta, M.~A. Kats, R.~Blanchard, G.~Aoust, J.-P.
  Tetienne, Z.~Gaburro, and F.~Capasso, ``Flat optics: controlling wavefronts
  with optical antenna metasurfaces,'' \emph{IEEE Journal of Selected Topics in
  Quantum Electronics}, vol.~19, no.~3, pp. 4\,700\,423--4\,700\,423, 2013.

\bibitem{qin2022metasurface}
J.~Qin, S.~Jiang, Z.~Wang, X.~Cheng, B.~Li, Y.~Shi, D.~P. Tsai, A.~Q. Liu,
  W.~Huang, and W.~Zhu, ``Metasurface micro/nano-optical sensors: principles
  and applications,'' \emph{ACS nano}, vol.~16, no.~8, pp. 11\,598--11\,618,
  2022.

\bibitem{kim_optical_2025}
H.~Kim, H.~Yun, S.~Jeong, S.~Lee, E.~Cho, and J.~Rho, ``Optical {Metasurfaces}
  for {Biomedical} {Imaging} and {Sensing},'' \emph{ACS Nano}, vol.~19, no.~3,
  pp. 3085--3114, Jan. 2025, publisher: American Chemical Society.

\bibitem{hu_3d-integrated_2019}
Y.~Hu, X.~Luo, Y.~Chen, Q.~Liu, X.~Li, Y.~Wang, N.~Liu, and H.~Duan,
  ``\BIBforeignlanguage{en}{{3D}-{Integrated} metasurfaces for full-colour
  holography},'' \emph{\BIBforeignlanguage{en}{Light: Science \&
  Applications}}, vol.~8, no.~1, p.~86, Sep. 2019, publisher: Nature Publishing
  Group.

\bibitem{nie_metasurfaces_2021}
S.~Nie and I.~F. Akyildiz, ``\BIBforeignlanguage{en}{Metasurfaces for
  multiplexed communication},'' \emph{\BIBforeignlanguage{en}{Nature
  Electronics}}, vol.~4, no.~3, pp. 177--178, Mar. 2021, publisher: Nature
  Publishing Group.

\bibitem{liu_programmable_2022}
C.~Liu, Q.~Ma, Z.~J. Luo, Q.~R. Hong, Q.~Xiao, H.~C. Zhang, L.~Miao, W.~M. Yu,
  Q.~Cheng, L.~Li, and T.~J. Cui, ``\BIBforeignlanguage{en}{A programmable
  diffractive deep neural network based on a digital-coding metasurface
  array},'' \emph{\BIBforeignlanguage{en}{Nature Electronics}}, vol.~5, no.~2,
  pp. 113--122, Feb. 2022, publisher: Nature Publishing Group.

\bibitem{teixeira_time-domain_2008}
F.~L. Teixeira, ``Time-{Domain} {Finite}-{Difference} and {Finite}-{Element}
  {Methods} for {Maxwell} {Equations} in {Complex} {Media},'' \emph{IEEE
  Transactions on Antennas and Propagation}, vol.~56, no.~8, pp. 2150--2166,
  Aug. 2008.

\bibitem{evanschitzky_fast_2007}
P.~Evanschitzky and A.~Erdmann, ``Fast near field simulation of optical and
  {EUV} masks using the waveguide method,'' in \emph{Proceedings of the SPIE,
  Volume 6533, article id. 65330Y}, vol. 6533, Feb. 2007, p. 65330Y, conference
  Name: 23rd European Mask and Lithography Conference ADS Bibcode:
  2007SPIE.6533E..0YE.

\bibitem{kang_large-scale_2024}
C.~Kang, C.~Park, M.~Lee, J.~Kang, M.~S. Jang, and H.~Chung, ``Large-scale
  photonic inverse design: computational challenges and breakthroughs,''
  \emph{Nanophotonics}, vol.~13, no.~20, pp. 3765--3792, 2024.

\bibitem{an_deep_2022}
S.~An, B.~Zheng, M.~Y. Shalaginov, H.~Tang, H.~Li, L.~Zhou, Y.~Dong,
  M.~Haerinia, A.~M. Agarwal, C.~Rivero-Baleine, M.~Kang, K.~A. Richardson,
  T.~Gu, J.~Hu, C.~Fowler, and H.~Zhang, ``\BIBforeignlanguage{en}{Deep
  {Convolutional} {Neural} {Networks} to {Predict} {Mutual} {Coupling}
  {Effects} in {Metasurfaces}},'' \emph{\BIBforeignlanguage{en}{Advanced
  Optical Materials}}, vol.~10, no.~3, p. 2102113, 2022.

\bibitem{li_pisc-net_2024}
C.~Li, J.~Chen, Q.~Lin, and Y.~Han, ``{PISC}-{Net}: {A} {Comprehensive}
  {Neural} {Network} {Framework} for {Predicting} {Metasurface} {Infrared}
  {Emission} {Spectra},'' \emph{ACS Applied Materials \& Interfaces}, vol.~16,
  no.~32, pp. 42\,816--42\,827, Aug. 2024, publisher: American Chemical
  Society.

\bibitem{cuomo_scientific_2022}
S.~Cuomo, V.~S. Di~Cola, F.~Giampaolo, G.~Rozza, M.~Raissi, and F.~Piccialli,
  ``\BIBforeignlanguage{en}{Scientific {Machine} {Learning} {Through}
  {Physics}–{Informed} {Neural} {Networks}: {Where} we are and {What}’s
  {Next}},'' \emph{\BIBforeignlanguage{en}{Journal of Scientific Computing}},
  vol.~92, no.~3, p.~88, Jul. 2022.

\bibitem{raissi_physics_2017}
M.~Raissi, P.~Perdikaris, and G.~E. Karniadakis,
  ``\BIBforeignlanguage{en}{Physics {Informed} {Deep} {Learning} ({Part} {II}):
  {Data}-driven {Discovery} of {Nonlinear} {Partial} {Differential}
  {Equations}},'' Nov. 2017, arXiv:1711.10566 [cs].

\bibitem{toscano_pinns_2025}
J.~D. Toscano, V.~Oommen, A.~J. Varghese, Z.~Zou, N.~Ahmadi~Daryakenari, C.~Wu,
  and G.~E. Karniadakis, ``\BIBforeignlanguage{en}{From {PINNs} to {PIKANs}:
  recent advances in physics-informed machine learning},''
  \emph{\BIBforeignlanguage{en}{Machine Learning for Computational Science and
  Engineering}}, vol.~1, no.~1, p.~15, Mar. 2025.

\bibitem{tu_physics-informed_2023}
J.~Tu, C.~Liu, and P.~Qi, ``Physics-{Informed} {Neural} {Network} {Integrating}
  {PointNet}-{Based} {Adaptive} {Refinement} for {Investigating} {Crack}
  {Propagation} in {Industrial} {Applications},'' \emph{IEEE Transactions on
  Industrial Informatics}, vol.~19, no.~2, pp. 2210--2218, Feb. 2023.

\bibitem{nguyen_physics-informed_2024}
D.~H. Nguyen, T.~H. Nguyen, K.~D. Tran, and K.~P. Tran,
  ``\BIBforeignlanguage{en}{Physics-{Informed} {Machine} {Learning} for
  {Industrial} {Reliability} and {Safety} {Engineering}: {A} {Review} and
  {Perspective}},'' in \emph{\BIBforeignlanguage{en}{Artificial {Intelligence}
  for {Safety} and {Reliability} {Engineering}: {Methods}, {Applications}, and
  {Challenges}}}, K.~P. Tran, Ed.\hskip 1em plus 0.5em minus 0.4em\relax Cham:
  Springer Nature Switzerland, 2024, pp. 5--23.

\bibitem{gopakumar_loss_2023}
V.~Gopakumar, S.~Pamela, and D.~Samaddar, ``\BIBforeignlanguage{en}{Loss
  landscape engineering via {Data} {Regulation} on {PINNs}},''
  \emph{\BIBforeignlanguage{en}{Machine Learning with Applications}}, vol.~12,
  p. 100464, Jun. 2023.

\bibitem{wang_experts_2023}
S.~Wang, S.~Sankaran, H.~Wang, and P.~Perdikaris, ``An {Expert}'s {Guide} to
  {Training} {Physics}-informed {Neural} {Networks},'' Aug. 2023,
  arXiv:2308.08468 [cs].

\bibitem{wang_understanding_2021}
S.~Wang, Y.~Teng, and P.~Perdikaris, ``Understanding and {Mitigating}
  {Gradient} {Flow} {Pathologies} in {Physics}-{Informed} {Neural}
  {Networks},'' \emph{SIAM Journal on Scientific Computing}, vol.~43, no.~5,
  pp. A3055--A3081, Jan. 2021, publisher: Society for Industrial and Applied
  Mathematics.

\bibitem{10.1063/5.0071616}
J.~Lim and D.~Psaltis, ``Maxwellnet: Physics-driven deep neural network
  training based on maxwell’s equations,'' \emph{APL Photonics}, vol.~7,
  no.~1, p. 011301, 01 2022.

\bibitem{10618982}
H.~Liu, Y.~Fan, F.~Ding, L.~Du, J.~Zhao, C.~Sun, and H.~Zhou,
  ``Physics-informed deep model for fast time-domain electromagnetic simulation
  and inversion,'' \emph{IEEE Transactions on Antennas and Propagation},
  vol.~72, no.~10, pp. 7807--7820, 2024.

\bibitem{Wang:24}
Y.~Wang and S.~Zhang, ``Multi-receptive-field physics-informed neural network
  for complex electromagnetic media,'' \emph{Opt. Mater. Express}, vol.~14,
  no.~11, pp. 2740--2754, Nov 2024.

\bibitem{kojima_inverse_2023}
K.~Kojima, T.~Koike‐Akino, Y.~Tang, and Y.~Wang, ``Inverse {Design} for
  {Integrated} {Photonics} {Using} {Deep} {Neural} {Network},'' in
  \emph{Integrated {Nanophotonics}: {Platforms}, {Devices}, and
  {Applications}}.\hskip 1em plus 0.5em minus 0.4em\relax Wiley, 2023, pp.
  209--243.

\bibitem{medvedev_3d_2024_1}
V.~Medvedev, A.~Erdmann, and A.~Rosskopf, ``{3D EUV mask simulator based on
  physics-informed neural networks: effects of polarization and
  illumination},'' in \emph{Computational Optics 2024}, vol. 13023,
  International Society for Optics and Photonics.\hskip 1em plus 0.5em minus
  0.4em\relax SPIE, 2024, p. 1302304.

\bibitem{medvedev_3d_2024}
V.~{M}edvedev, A.~Erdmann, and A.~Rosskopf, ``{3D} mask simulation and
  lithographic imaging using physics-informed neural networks,'' in
  \emph{Optical and {EUV} {Nanolithography} {XXXVII}}, vol. 12953.\hskip 1em
  plus 0.5em minus 0.4em\relax SPIE, Apr. 2024, pp. 208--224.

\bibitem{Medvedev:25}
V.~Medvedev, A.~Erdmann, and A.~Rosskopf, ``Physics-informed deep learning for
  {3D} modeling of light diffraction from optical metasurfaces,'' \emph{Opt.
  Express}, vol.~33, no.~1, pp. 1371--1384, Jan 2025.

\bibitem{graham_helmholtz_2019}
I.~Graham, O.~Pembery, and E.~Spence, ``\BIBforeignlanguage{en}{The {Helmholtz}
  equation in heterogeneous media: {A} priori bounds, well-posedness, and
  resonances},'' \emph{\BIBforeignlanguage{en}{Journal of Differential
  Equations}}, vol. 266, no.~6, pp. 2869--2923, Mar. 2019.

\bibitem{lim_maxwellnet_2022}
J.~Lim and D.~Psaltis, ``{MaxwellNet}: {Physics}-driven deep neural network
  training based on {Maxwell}’s equations,'' \emph{APL Photonics}, vol.~7,
  no.~1, p. 011301, Jan. 2022.

\bibitem{garcia-garcia_pointnet_2016}
A.~Garcia-Garcia, F.~Gomez-Donoso, J.~Garcia-Rodriguez, S.~Orts-Escolano,
  M.~Cazorla, and J.~Azorin-Lopez, ``{PointNet}: {A} {3D} {Convolutional}
  {Neural} {Network} for real-time object class recognition,'' in \emph{2016
  {International} {Joint} {Conference} on {Neural} {Networks} ({IJCNN})}, Jul.
  2016, pp. 1578--1584, iSSN: 2161-4407.

\bibitem{qi_pointnet_2017}
C.~R. Qi, L.~Yi, H.~Su, and L.~J. Guibas, ``{PointNet}++: {Deep} {Hierarchical}
  {Feature} {Learning} on {Point} {Sets} in a {Metric} {Space},'' in
  \emph{Advances in {Neural} {Information} {Processing} {Systems}},
  vol.~30.\hskip 1em plus 0.5em minus 0.4em\relax Curran Associates, Inc.,
  2017.

\bibitem{kashefi_physics-informed_2022}
A.~Kashefi and T.~Mukerji, ``\BIBforeignlanguage{en}{Physics-informed
  {PointNet}: {A} deep learning solver for steady-state incompressible flows
  and thermal fields on multiple sets of irregular geometries},''
  \emph{\BIBforeignlanguage{en}{Journal of Computational Physics}}, vol. 468,
  p. 111510, Nov. 2022.

\bibitem{kashefi_physics-informed_2023}
A.~Kashefi, L.~J. Guibas, and T.~Mukerji,
  ``\BIBforeignlanguage{en}{Physics-informed {PointNet}: {On} how many
  irregular geometries can it solve an inverse problem simultaneously?
  {Application} to linear elasticity},'' Sep. 2023, arXiv:2303.13634 [cs].

\bibitem{kashefi_kolmogorov-arnold_2024}
A.~Kashefi and T.~Mukerji, ``Physics-informed pointnet: A deep learning solver
  for steady-state incompressible flows and thermal fields on multiple sets of
  irregular geometries,'' \emph{J. Comput. Phys.}, vol. 468, p. 111510, 2022.

\bibitem{liu_kan_2024}
Z.~Liu, Y.~Wang, S.~Vaidya, F.~Ruehle, J.~Halverson, M.~Soljačić, T.~Y. Hou,
  and M.~Tegmark, ``\BIBforeignlanguage{en}{{KAN}: {Kolmogorov}-{Arnold}
  {Networks}},'' Jun. 2024, arXiv:2404.19756 [cond-mat, stat].

\bibitem{ZHANG2022110179}
Z.~Zhang, ``A physics-informed deep convolutional neural network for simulating
  and predicting transient darcy flows in heterogeneous reservoirs without
  labeled data,'' \emph{Journal of Petroleum Science and Engineering}, vol.
  211, p. 110179, 2022.

\bibitem{kingma_adam_2017}
D.~P. Kingma and J.~Ba, ``Adam: {A} {Method} for {Stochastic} {Optimization},''
  Jan. 2017, arXiv:1412.6980 [cs].

\bibitem{kashefi_prediction_2023}
A.~Kashefi and T.~Mukerji, ``Prediction of fluid flow in porous media by sparse
  observations and physics-informed {PointNet},'' \emph{Neural Networks}, vol.
  167, pp. 80--91, Oct. 2023.

\bibitem{glorot_understanding_2010}
X.~Glorot and Y.~Bengio, ``\BIBforeignlanguage{en}{Understanding the difficulty
  of training deep feedforward neural networks},'' in
  \emph{\BIBforeignlanguage{en}{Proceedings of the {Thirteenth} {International}
  {Conference} on {Artificial} {Intelligence} and {Statistics}}}.\hskip 1em
  plus 0.5em minus 0.4em\relax JMLR Workshop and Conference Proceedings, Mar.
  2010, pp. 249--256, iSSN: 1938-7228.

\bibitem{pmlr-v97-rahaman19a}
N.~Rahaman, A.~Baratin, D.~Arpit, F.~Draxler, M.~Lin, F.~Hamprecht, Y.~Bengio,
  and A.~Courville, ``On the spectral bias of neural networks,'' in
  \emph{Proceedings of the 36th International Conference on Machine Learning},
  ser. Proceedings of Machine Learning Research, K.~Chaudhuri and
  R.~Salakhutdinov, Eds., vol.~97.\hskip 1em plus 0.5em minus 0.4em\relax PMLR,
  09--15 Jun 2019, pp. 5301--5310.

\bibitem{wu_comprehensive_2023}
C.~Wu, M.~Zhu, Q.~Tan, Y.~Kartha, and L.~Lu, ``\BIBforeignlanguage{en}{A
  comprehensive study of non-adaptive and residual-based adaptive sampling for
  physics-informed neural networks},'' \emph{\BIBforeignlanguage{en}{Computer
  Methods in Applied Mechanics and Engineering}}, vol. 403, p. 115671, Jan.
  2023.

\bibitem{imran_fourierkan_2024}
A.~A. Imran and M.~F. Ishmam, ``{FourierKAN} outperforms {MLP} on {Text}
  {Classification} {Head} {Fine}-tuning,'' Sep. 2024, arXiv:2408.08803 [cs].

\bibitem{li_kolmogorov-arnold_2024}
Z.~Li, ``Kolmogorov-{Arnold} {Networks} are {Radial} {Basis} {Function}
  {Networks},'' May 2024, arXiv:2405.06721 [cs].

\bibitem{medvedev_physics-informed_2025}
V.~Medvedev, A.~Erdmann, and A.~Rosskopf,
  ``\BIBforeignlanguage{en}{Physics-informed deep learning for {3D} modeling of
  light diffraction from optical metasurfaces},''
  \emph{\BIBforeignlanguage{en}{Optics Express}}, vol.~33, no.~1, p. 1371, Jan.
  2025.

\end{thebibliography}
\addcontentsline{toc}{section}{Bibliography}

\begin{IEEEbiography}[{\includegraphics[width=1in,height=1.25in,clip,keepaspectratio]{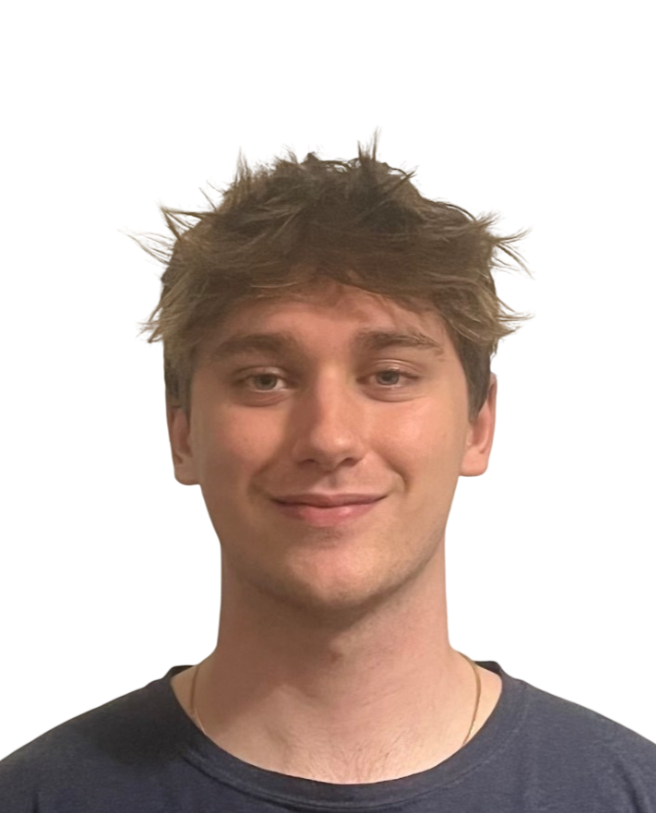}}]{Leon Armbruster}
received his MSc in Applied and Computational Mathematics from Friedrich Alexander-Universität, Erlangen, Germany, in 2025. He is currently working in the AI-Augmented Simulation research group at the Fraunhofer IISB in Erlangen, Germany. 
\end{IEEEbiography}

\begin{IEEEbiography}[{\includegraphics[width=1in,height=1.25in,clip,keepaspectratio]{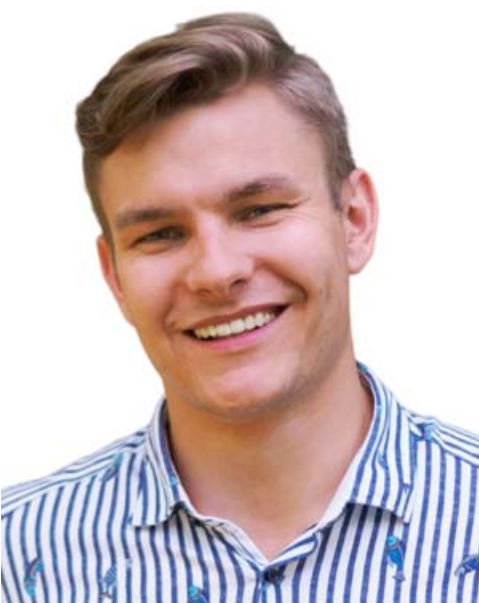}}]{Vlad Medvedev}
received his MSc in Advanced Optical Technologies from Friedrich Alexander-Universität, Erlangen, Germany, in 2022. He is currently working towards a PhD in physics-informed deep learning for computational lithography and design of nanooptical devices with the Computational Lithography and AI-Augmented Simulation research groups at the Fraunhofer IISB in Erlangen, Germany. His research interests include the implementation of computer vision and deep learning solutions to enhance lithographic output.
\end{IEEEbiography}
\begin{IEEEbiography}[{\includegraphics[width=1in,height=1.25in,clip,keepaspectratio]{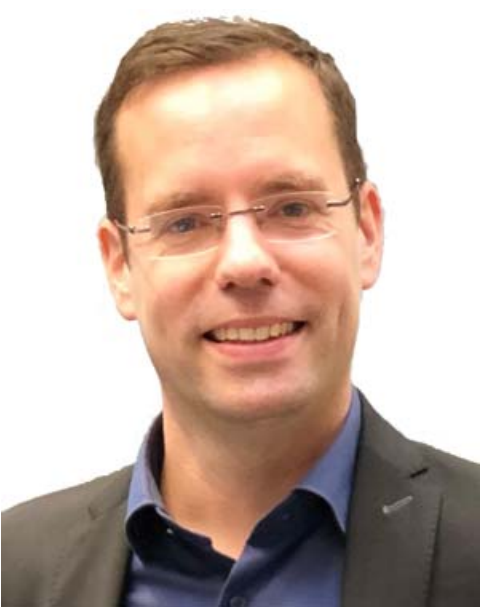}}]{Dr. Andreas Roßkopf}
received his Diploma in Applied Mathematics and the PhD in Engineering from Friedrich-Alexander-Universität, Erlangen, Germany, in 2008 and 2018, respectively. He has worked in numerous positions in the automotive, mobility, and energy industries. Since 2018, he has been leading the research group AI-Augmented Simulation at the Fraunhofer IISB in Erlangen, Germany. His research interests include coupled numerical simulation, design automation, and optimization of engineering processes through data and AI.
\end{IEEEbiography}

\end{document}